\documentclass[12pt,letterpaper]{article}
\usepackage[a4paper, total={7in, 10in}]{geometry}

\usepackage{graphicx}
\usepackage{helvet}
\usepackage{authblk}
\usepackage{hyperref}
\usepackage{amsmath} 
\usepackage{amssymb} 
\usepackage{orcidlink} 
\usepackage[super,comma,sort&compress]  
   {natbib}

\usepackage{pdflscape}

\usepackage{multirow}
\usepackage{array}
\usepackage{bm} 
\usepackage{multirow}
\usepackage{tabularx}
\usepackage{booktabs}
\newcolumntype{P}[1]{>{\centering\arraybackslash}p{#1}}
\newcolumntype{M}[1]{>{\centering\arraybackslash}m{#1}}
\usepackage{caption}
\usepackage{longtable}
\usepackage{float}

\usepackage{subcaption}

\makeatletter
\renewcommand{\maketitle}{\bgroup\setlength{\parindent}{0pt}
\begin{flushleft}
  \textbf{\@title}
  
  \@author
\end{flushleft}\egroup}
\makeatother

\title{A Framework for Evaluating the Siting of Fusion Power: Case Study on the Retired Coal Sites in the United States}
\date{}

\author[1,2,*\orcidlink{0000-0003-4680-4875}]{Muhammad R. Abdussami}
\author[1,2]{Kevin Daley}
\author[1,2]{Gabrielle Hoelzle}
\author[1,2,3,**]{Aditi Verma}

\affil[1]{Department of Nuclear Engineering and Radiological Sciences, University of Michigan, 2355 Bonisteel Blvd, Ann Arbor, 48105, MI, United States}
\affil[2]{Fastest Path to Zero Initiative, University of Michigan, 2355 Bonisteel Blvd, Ann Arbor, 48105, MI, United States}
\affil[3]{Lead contact}

\affil[*]{Correspondence: rafiul@umich.edu}
\affil[**]{Correspondence: aditive@umich.edu }

\begin{document}

\maketitle

\section*{SUMMARY}





As fusion advances toward commercialization, systematic siting approaches are needed to identify locations that meet technical, economic, and infrastructural requirements, while also ensuring public acceptance and avoiding the socio-political challenges that have historically hindered fission deployment. Therefore, this study introduces a comprehensive, first-of-its-kind fusion siting framework and applies it to 85 retired (2020-2025) U.S. coal power sites as a case study. The framework evaluates 21 sub-criteria under four key attributes: State Policies, Federal Policies, Risk and Hazard Metrics, and Connectivity and Spatial Factors. Sub-attributes weights are derived using the “Fuzzy Full Consistency Method” with input from five fusion experts, and site rankings are determined using the “Measurement Alternatives and Ranking According to COmpromise Solution” method. Results indicate that federal incentives, transportation, substation, and energy prices are the most important factors for fusion siting. Sensitivity analysis reveals that landslide hazards have the greatest effect on rank stability, while fault lines is the least influential. A separate comparative assessment of the fusion deployment sites proposed by Type One Energy, Zap Energy, and Commonwealth Fusion Systems is also conducted using results from our proposed framework. This framework provides a transparent, stakeholder-inclusive decision-making tool that clarifies how sites are evaluated using weighted criteria and distinguishes inflexible policy-responsive factors, thereby enabling targeted regional and federal strategies.

\section*{KEYWORDS}


Fusion Power Plants, Siting, Coal Power Plants, Multi-Criteria Decision-Making (MCDM) methods 

\section{INTRODUCTION}


In recent years, fusion energy has transitioned from an aspirational research pursuit to a rapidly maturing global industry driven by technological breakthroughs and private-sector momentum. As of 2025, over 53 private fusion companies are operating across more than a dozen countries, marking a fourfold increase since 2021~\cite{GlobalFusionReport2025}. Collectively, these firms have attracted over \$9.7 billion in total investment, with \$2.6 billion raised in the past year alone~\cite{GlobalFusionReport2025}, reflecting unprecedented investor confidence and technological progress. Companies such as Commonwealth Fusion Systems (CFS), Helion Energy, TAE Technologies, General Fusion, Focused Energy, and Type One Energy have demonstrated significant milestones—ranging from high-field superconducting magnets and direct-energy-conversion prototypes to commercial power purchase agreements (PPAs) with major corporations like Microsoft, Google, and Nucor, signaling the industry’s first commercial commitments. Fusion startups are pursuing a diverse range of technological pathways, including magnetic confinement (tokamaks, stellarators), inertial confinement, magneto-inertial confinement, proton–boron fusion, and muon-catalyzed approaches, each contributing to a rich innovation ecosystem. Projections suggest that the majority of companies anticipate operating commercially viable pilot plants between 2030 and 2035, with grid-connected electricity expected shortly thereafter~\cite{GlobalFusionReport2025}. Together, these developments mark a turning point in fusion history—transforming it from a perpetual promise into an emerging clean-energy reality poised for commercialization within the next decade. While the technical readiness of fusion technologies is improving, identifying commercially and infrastructurally suitable pathways for deployment remains a critical challenge. Historically, government-led fusion programs have focused on scientific milestones rather than market readiness, resulting in devices “not well-suited for commercial application” due to their size, cost, and complexity \cite{pearson2020technology}.

Strategic siting decisions are critical not only for determining the technical suitability of fusion power projects—such as grid connectivity, cooling water access, seismic stability, and land availability—but also for shaping public acceptance, regulatory approval, and long-term infrastructure costs~\cite{mohamed2024global}. The history of fission energy development offers important lessons in this regard. During the mid-20th century, many nuclear power plants were sited primarily based on engineering and economic factors, often with limited public consultation or transparency~\cite{he2013public}. This technocratic approach contributed to social opposition, mistrust, and delayed projects in subsequent decades, as seen in siting conflicts over facilities such as Seabrook in the U.S.~\cite{o1983facility}. The failure of top-down siting approaches for spent-fuel repositories—exemplified by the halted Yucca Mountain Project—further underscores the importance of integrating community engagement, environmental justice, and equitable decision-making into early siting frameworks~\cite{richter2022process}. For emerging fusion technologies, learning from these fission-era missteps is essential to ensure that the next generation of clean-energy sites are not only technically sound but also socially legitimate and publicly trusted.

One promising strategy involves repurposing existing energy infrastructure, particularly retiring coal-fired power plants, as potential sites for fusion deployment. These sites offer valuable assets, including existing grid interconnections, water access, a permitting history, and a skilled local workforce~\cite{hansen2022investigating}. Notably, several public and private fusion developers are actively exploring this approach. For example, Zap Energy has launched a feasibility study, supported by the Centralia Coal Transition Board, to convert the TransAlta coal plant in Washington into a pilot fusion facility \cite{FusionZAPCoal}. Similarly, Type One Energy plans to build a stellarator fusion prototype at the former Bull Run coal plant in Tennessee  \cite{FusionTVACoal}, and CFS is evaluating former coal and gas plants for its first commercial fusion system \cite{FusionCFSCoal}. In the UK, the government has selected the West Burton coal plant as the site for its Spherical Tokamak for Energy Production (STEP) prototype \cite{FusionUKCoal}. 

The timelines for retiring coal-fired power plants in the U.S.—with 68,789 MW planned for retirement or conversion between 2025 and 2030~\cite{IEEFA2024}—align closely with the expected commercialization of fusion energy, as 35 out of 45 fusion companies aim to operate pilot plants between 2030 and 2035~\cite{EENews}, creating a strategic opportunity for site reuse. The Global Energy Monitor shows that in 2024, 25.2 GW of coal capacity was retired globally, and that many plants still lack a planned shutdown date, indicating a gradual transition~\cite{GEM2025}. In some major economies, the phase-out years are explicit: e.g., the G7 nations committed to ending the use of unabated coal power during the first half of the 2030s~\cite{Reuters2024}. This overlaps with the projected deployment window for first-of-a-kind (FOAK) and n-th-of-a-kind (NOAK) fusion facilities, which are expected to be in reality by 2030 to 2050 \cite{carayannis2021reviewing, katoch2025fusion}.
As a result, coal plant sites may serve as strategic launchpads for both early demonstration and long-term commercial fusion energy projects.

While existing research on coal-to-nuclear transitions has largely centered on fission reactors, the siting of fusion facilities remains underexplored. Given fusion's distinct advantages, such as an inherent safety profile~\cite{takeda2018nuclear}, absence of long-lived radioactive waste~\cite{de2022overview}, and fuel abundance~\cite{bradshaw2011nuclear}, there is a compelling need to develop siting methodologies tailored specifically to fusion \cite{mishra2020nuclear, ITER}. Approaches to fusion siting may also differ from those used to site fission facilities due to differences in public sentiment \cite{kawamoto2025public}. For example, a community that opposes fission due to nuclear waste concerns can be more open to fusion as it avoids the social and regulatory hurdles associated with spent fuel disposal \cite{PubPercFus}. Fission siting frameworks are primarily designed around managing risks from high-level radioactive waste, large exclusion zones, and continuous fuel-cycle logistics~\cite{eidelpes2022fission}—concerns that are minimal or absent in fusion systems. Similarly, coal-plant siting has historically optimized for proximity to fuel supply chains, rail access, and water availability rather than for electromagnetic confinement infrastructure, plasma containment, or advanced grid integration. Fusion facilities instead demand high electrical interconnection capacity, precision cooling, and minimal seismic interference while offering far smaller offsite radiological risk profiles and greater siting flexibility


To date, fusion research facilities have primarily been sited based on access to extensive scientific infrastructure—including high-power experimental halls, superconducting magnets and cryogenic laboratories, diagnostic facilities, and specialized human capital—along with opportunities for international collaboration and established research ecosystems~\cite{ITER2023}. For instance, Medrano et al. considered land, geological characteristics, freshwater supply, sanitary and industrial sewage, heat sink, electrical power supply, and transport and shipping for siting ITER in Spain \cite{medrano2003iter}. However, as fusion transitions from research to commercialization, regulatory landscapes are beginning to shift. In the U.S., the Nuclear Regulatory Commission (NRC) has recently clarified that fusion facilities will not be regulated under the same framework as fission reactors, but rather through a more flexible byproduct materials framework under 10 CFR Part 30 \cite{FusionNRC2023}. This regulatory distinction has significant siting implications, potentially streamlining site approval timelines and opening new opportunities. As fusion development moves toward FOAK deployments, attention is turning to brownfield sites, including retired coal plants, which may offer valuable infrastructure and community readiness for hosting future fusion facilities.

Screening and site selection tools play a crucial role in identifying locations that are suitable for powering with new energy technologies. Using the Oak Ridge Siting Analysis for Power Generation Expansion (OR-SAGE) tool, a study evaluates 34 coal plant sites across the U.S., incorporating geographic and environmental factors such as population density, water availability, and seismic activity. The results show that 77\% of these sites are viable for hosting fission Small Modular Reactors (SMRs), representing 7.3 GWe of coal-fired capacity that could be transitioned to nuclear energy \cite{belles2013evaluation,omitaomu2022methods}. Complementing this spatial approach, optimization models such as Mixed-Integer Nonlinear Programming (MINLP) have been applied to quantitatively assess the best sites for greenfield SMR deployment, balancing factors including cost, safety, and geographic suitability. While this method improves upon traditional qualitative siting strategies, it is constrained by the relatively limited number of attributes considered in the analysis \cite{devanand2019optimal}. Broader feasibility assessments have incorporated spatial and policy-based screening tools such as the Siting Tool for Advanced Nuclear Development (STAND), which can evaluate fission-based nuclear reactor siting based on safety, socioeconomic, and regulatory conditions \cite{abdussami2024investigation}. However, STAND fails to incorporate the perspective of multiple decision-makers.

On the contrary, Multi-criteria decision-making (MCDM) frameworks have been extensively applied in siting decisions for both nuclear and non-nuclear facilities because they provide a structured approach to balance complex, often conflicting criteria—such as safety, environmental impact, socio-economic benefits, and technical feasibility—under conditions of uncertainty. In such contexts, quantitative data alone are often insufficient, as many attributes (e.g., perceived risk, social acceptance, political stability, or long-term safety assurance) cannot be directly measured. Expert judgment, therefore, becomes essential for assigning realistic weights to these qualitative factors, integrating experiential knowledge, regulatory insight, and contextual understanding into the decision process. Applications of MCDM techniques span a wide range of non-nuclear contexts, including landfill siting in Iraq \cite{alkaradaghi2019landfill}, offshore wind farm selection in Iran \cite{fetanat2015novel}, solar power plant selection in Indonesia \cite{wang2023site}, shopping center site selection in Turkey \cite{onut2010combined}, and urban hospital location selection in Iran \cite{alavi2013proper}. Across these studies, the intended audiences are primarily urban planners, policymakers, investors, and infrastructure developers who must select locations that balance technical feasibility, environmental sustainability, social acceptability, and economic efficiency. A common theme uniting these applications is the use of MCDM frameworks to conduct systematic, transparent, and reproducible analyses that integrate both quantitative data (e.g., land use, resource availability, transport networks) and qualitative expert judgment (e.g., social impact, safety, and regulatory feasibility).

A wide range of fuzzy and hybrid MCDM approaches—including Fuzzy-AHP, Fuzzy-TOPSIS, Z-number-based methods (e.g., Z-BWM, Z-DEMATEL), and GIS-integrated AHP models—have been employed in nuclear siting studies to handle uncertainty, interdependence, and expert judgment across diverse technical, environmental, and socio-economic criteria \cite{shafii2024prioritizing,peng2020multi,erol2014fuzzy,susiati2022site,abdullah2023multi,topalouglu2025analytic,erdougan2016combined,mahmudah2024developing}.
These methods collectively highlight the importance of structured, expert-driven decision tools for balancing multi-dimensional risks and benefits in reactor siting.
Although these studies identify the relative importance of siting factors, none have applied these significance values/weights across a broad set of candidate sites or practical case studies to determine the most suitable location.

Repurposing retiring coal power plants for fission nuclear deployment has been shown to be technically and economically feasible in a wide range of contexts. Studies estimate that up to 80\% of evaluated coal sites in the U.S. could accommodate advanced reactors based on siting characteristics, cost considerations, and regional economic impacts \cite{hansen2022investigating}. Recent studies across South Korea, China, and Poland demonstrate that repowering coal plants with advanced or SMRs offers significant technical, economic, and environmental advantages—including grid compatibility, enhanced regional resilience, and optimized use of existing water and steam infrastructure \cite{joo2024evaluation, weng2024impact, ochmann2024potential}. Ensuring legal and safety compliance in coal-to-nuclear conversions extends beyond meeting reactor design codes—it involves comprehensive site evaluation, licensing alignment, and adherence to international nuclear safety and environmental protection standards. In the siting context, this includes early integration of regulatory review processes, environmental impact assessments, seismic and hydrological analyses, and public consultation mechanisms consistent with IAEA safety standards~\cite{chmielewska2024selected}. Literature shows that robust siting and early stakeholder engagement can substantially reduce project delays, regulatory conflicts, and local opposition, where transparent site-selection frameworks accelerated licensing and improved community acceptance~\cite{stanitsas2024navigating, susskind2022sources}. Retired coal plant sites offer ready grid connections, cooling systems, and industrial zoning~\cite{hansen2022investigating}, making them ideal for fusion plant deployment. These attributes, already proven advantageous for coal-to-fission conversions, can similarly support cost-effective fusion siting.


While numerous studies have explored methods for siting fission-based nuclear plants, none have developed a framework for siting fusion reactors. Given the distinct siting requirements of fusion power, such as high electrical interconnection capacity, stable seismic and geological conditions, access to abundant cooling water, availability of skilled technical labor, and proximity to existing industrial or research infrastructure, relying on fission siting methodologies and the criteria weights for fusion site selection is inappropriate. Additionally, prior research often considers a limited set of criteria for fission reactor siting, whereas real-world fission or fusion reactor siting demands a more holistic evaluation of diverse environmental, economic, and technical factors, such as seismic risk, cooling water access, land and population constraints, grid connectivity, environmental impact, and community acceptance. To address these research gaps, this study proposes a first-of-its-kind framework for siting fusion reactors, specifically designed to capture the distinct requirements of fusion technology. We demonstrate this framework by applying it to a comprehensive case study of repurposing 85  retired coal plant sites across the U.S, excluding Alaska and Hawaii due to data unavailability. The proposed methodology incorporates 21 carefully selected siting criteria covering environmental, economic, and technical dimensions to enable a rigorous assessment of candidate sites. To ensure methodological rigor, we employ two advanced multi-criteria decision-making techniques, Fuzzy- Full Consistency Method (F-FUCOM) for deriving robust criteria weights and Measurement Alternatives and Ranking According to COmpromise Solution (MARCOS) for systematically ranking the suitability of coal sites for fusion deployment. 

Furthermore, we conduct a sensitivity analysis to evaluate the impact of variations in criteria weights and rank reversals. This study also provides one of the first granular, multi-attribute evaluations of the proposed sites of leading U.S. fusion developers—Type One Energy’s Bull Run, Zap Energy’s Centralia, and Commonwealth Fusion Systems’ Chesterfield—revealing how sub–attribute–level factors drive their relative suitability.

\section{METHODS}
\label{sec: methods}

\subsection{Modeling approach overview}

\begin{figure}[htbp]
\centering
\includegraphics[width=0.9\textwidth, height=0.9\textheight, keepaspectratio]{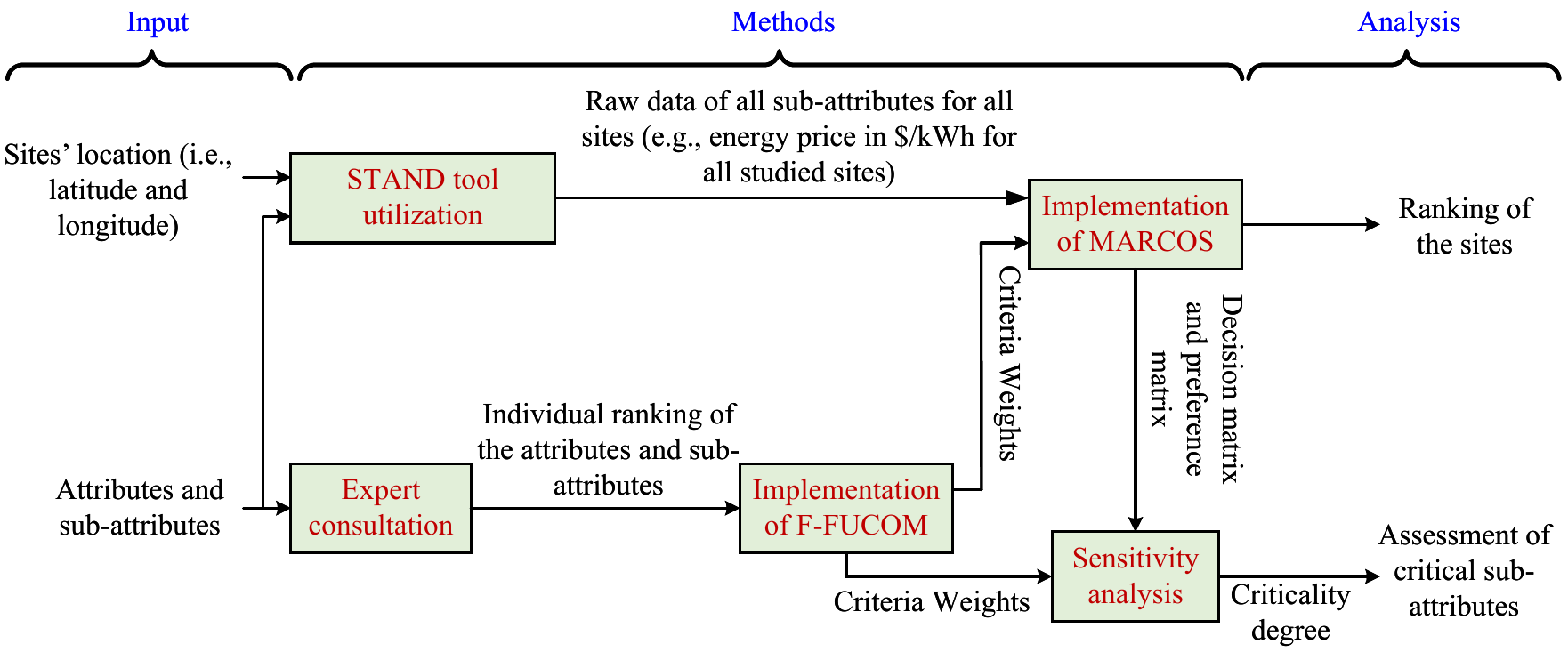}

\caption{Flowchart of the research methodology.}
\label{fig: Research Methodology}
\end{figure}

The research methodology is outlined in Figure \ref{fig: Research Methodology}, which presents a structured approach for evaluating the suitability of fusion siting. In the initial stage, we classify the siting criteria into four primary categories/attributes: State Policies (SP), Federal Policies (FP), Risk and Hazard Metrics (RHM), and Connectivity and Spatial Factors (CSF). Each category is further subdivided into specific attributes. For the case study, we use data from 85 retired coal-fired power plants across the U.S. between 2020 and 2025 \cite{EAIPowerPlants}. However, due to data unavailability for Alaska and Hawaii, two retired coal plant sites from these regions between 2020 and 2025 are excluded. The dataset of 85 coal sites is provided in the Code and data availability section.

The geographic coordinates (latitude and longitude) of each site, along with the identified siting attributes, are then input into the Siting Tool for Advanced Nuclear Development (STAND) to extract raw data for all relevant sub-attributes \cite{FPZ}. For instance, net electricity import, a representative sub-attribute, is retrieved from STAND in million kWh for each coal site. Subsequently, expert consultations are conducted to gather individual rankings of the attributes and sub-attributes from a panel of Fusion Experts (FEs). These expert rankings are then processed using the Fuzzy Full Consistency Method (F-FUCOM) to derive the weight of each attribute and sub-attribute. The resulting criteria weights, along with the raw data from STAND, serve as inputs to the Measurement of Alternatives and Ranking according to COmpromise Solution (MARCOS) method, which is used to rank the coal sites. Finally, a comprehensive sensitivity analysis is performed using the F-FUCOM-derived weights, as well as the decision and preference matrices from MARCOS. The following sections provide detailed explanations of each methodological step.

\subsection{Utilization of the STAND tool}

\begin{figure}[htbp]
\centering
\includegraphics[width=0.9\textwidth, height=0.8\textheight, keepaspectratio]{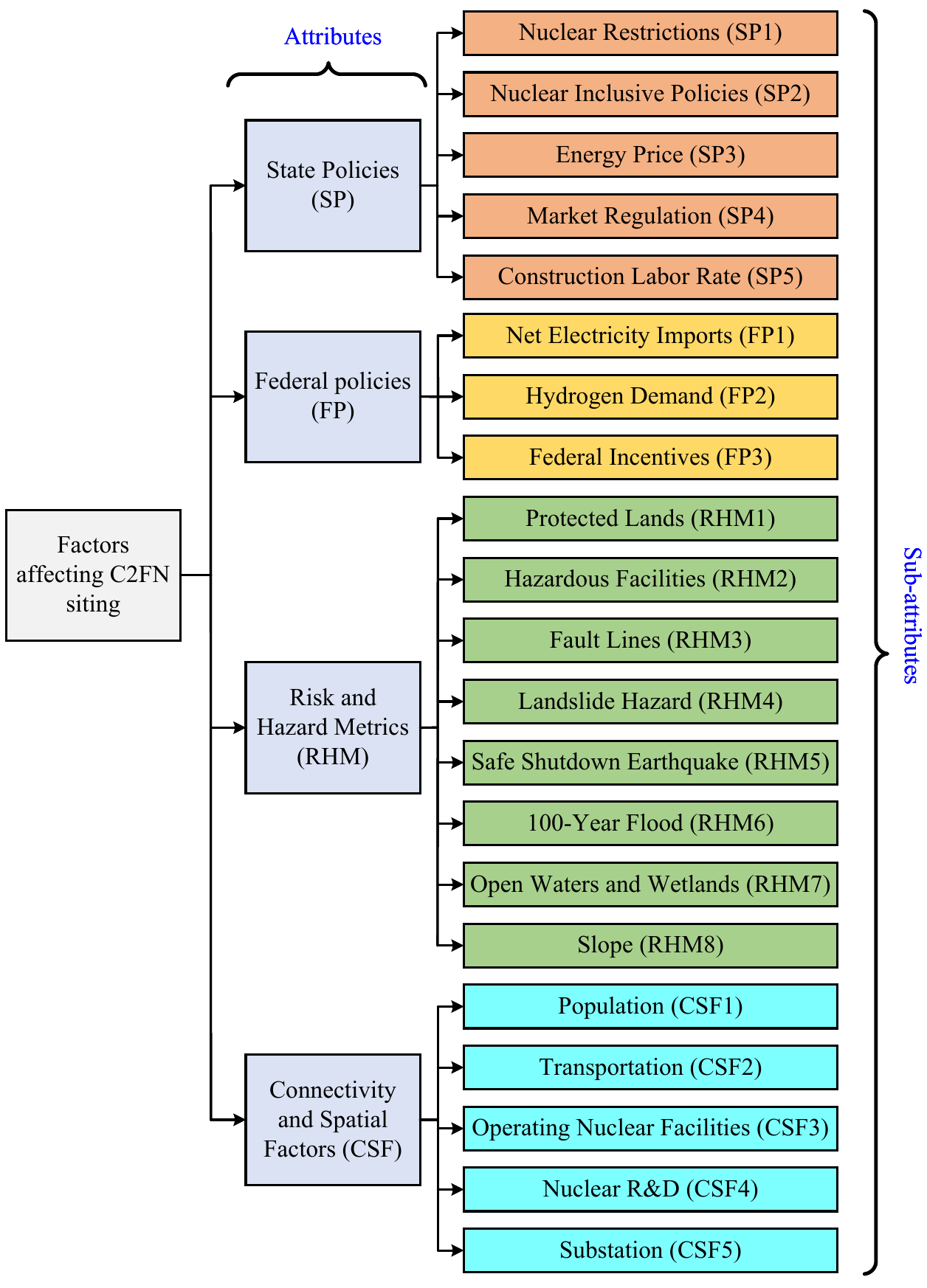}
\caption{List of attributes and sub-attributes for fusion siting study.}\label{fig: list of attribute}
\end{figure}

The STAND takes input as the location of the sites and their sub-attributes and provides the raw data for these sub-attributes. The complete list of attributes and their corresponding data is provided in the Code and data availability section. Several attributes in the dataset contain binary values represented as True and False (e.g., fault line hazard presence). These text-based values are converted into numerical representations to facilitate computational processing, where True = 1 and False = 0. 
The identified attributes and sub-attributes are presented in Figure \ref{fig: list of attribute}. A brief discussion on each of the sub-attributes and the rationale is provided in supplemental information (see Note S1).

\subsection{Consultations with fusion experts (FEs)}

We conduct interview sessions with five FEs and distributed an online survey to gather their input on each identified attribute. The survey utilized a 1-5 scale, ranging from “Least Important” to “Extremely Important,” to assess the significance of each attribute. A summary of the participating FEs is presented in Table \ref{tab: Details of the Fusion Experts}.

\begin{table}[htbp]
\centering
\small
\caption{Details of the Fusion Experts.}
\label{tab: Details of the Fusion Experts}
\begin{tabular}{|M{3.0cm}|M{2.8cm}|M{4.6cm}|M{2cm}|}
\hline
\textbf{Affiliated organization} & \textbf{Designation} & \textbf{Educational qualification (highest degree)} & 
\textbf{Experience (years)} \\
\hline
Private & Vice President  & Masters & 6+  \\
\hline

Private & Vice President  & Ph.D.   & 7+  \\
\hline

Government     & Advisor         & Masters & 5+  \\
\hline

Government     & Senior Advisor  & Ph.D.   & 25+ \\
\hline

Private & Founder         & Masters & 8+  \\

\hline
\end{tabular}
\label{tab:fusion_experts}
\end{table}

\subsection{Implementation of F-FUCOM}

We employ an advanced MCDM method, F-FUCOM, to determine the weights of the attributes. Unlike traditional FUCOM, which relies on crisp pairwise comparisons, F-FUCOM integrates linguistic scales and triangular fuzzy numbers to better capture expert uncertainty and subjectivity \cite{pamucar2020prioritizing}. F-FUCOM ensures full consistency in weight derivation and reduces inconsistencies typical in the AHP and Best-Worst Method (BWM). By incorporating fuzzy logic, F-FUCOM allows experts to express preferences using linguistic terms, addressing vagueness and imprecision, which is not effectively handled in traditional FUCOM. Unlike AHP and BWM, which require n(n-1)/2 and (2n-3) pairwise comparisons, respectively, FUCOM-F requires (n-1) pairwise comparisons, where n is the number of criteria \cite{ilieva2020fuzzy}. This makes F-FUCOM more efficient for large-scale problems. F-FUCOM minimizes the impact of subjective bias by using a structured fuzzy decision framework. Although F-FUCOM is a relatively recent MCDM method, researchers use it in several applications, such as evaluating the sustainability of farm tourism sites \cite{ocampo2022full}, transportation demand management \cite{pamucar2020fuzzy}, and telecommunication quality improvements \cite{ccodur2024application}. The detailed implementation procedure of F-FUCOM consists of four key steps. First, we identify and define the evaluation criteria for the coal-to-fusion siting problem. We have a total of four criteria, and each criterion comprises several sub-criteria, as depicted in Figure~\ref {fig: list of attribute}. For n number of criteria, the criteria set can be defined by Equation~\ref{eq: FUCOM1}. 
\begin{equation}
{C} = \{C_1, C_2, \ldots, C_n\}
\label{eq: FUCOM1}
\end{equation}

Second, FEs rank the criteria based on their significance. The most important criterion receives the highest rank, while the least important gets the lowest rank, as presented in Equation~\ref{eq: FUCOM2}.
\begin{equation}
C_{j(1)} > C_{j(2)} > \cdots > C_{j(k)}
\label{eq: FUCOM2}
\end{equation}

where k is the rank of the observed criteria. If two or more criteria have the same ranking, an equality sign is placed between them instead of a “greater than” sign.

Then, the criteria are compared against the most significant criterion using predefined fuzzy linguistic terms. The fuzzy linguistic scale used in this research is presented in Table \ref{tab: Fuzzy scale} \cite{soltani2011hospital}.

\begin{table}[htbp]
\centering
\small
\caption{Fuzzy scale.}
\label{tab: Fuzzy scale}
\begin{tabular}{|c|c|}
\hline
\textbf{Linguistic Terms} & \textbf{Membership Function} \\
\hline
Equally Significant (ES)         & (1,1,1) \\
\hline

Moderately Significant (MS)      & (2,3,4) \\
\hline

Strongly Significant (SS)        & (4,5,6) \\
\hline

Very Strongly Significant (VSS)  & (6,7,8) \\
\hline

Extremely Significant (ExS)      & (9,9,9) \\
\hline
\end{tabular}
\label{tab:fuzzy_scale}
\end{table}

\begin{figure}[htbp]
\centering
\includegraphics[width=0.5\textwidth, height=0.5\textheight, keepaspectratio]{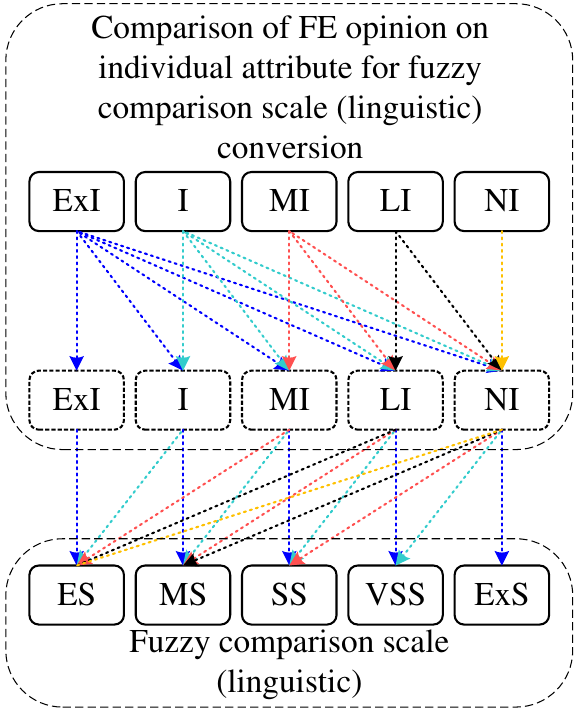}
\caption{Methodology to develop fuzzy comparison scale (linguistic) from FE opinion.}
\label{fig: scale_conversion}
\end{figure}

Each criterion $C_k$ is compared against the most significant criterion $C_{j(1)}$. Since the highest-ranked criterion is always compared with itself, its significance is categorized as ES. Since FE interviews are conducted using a distinct linguistic scale, a scale conversion process is necessary to align their opinions with the fuzzy criteria significance $\left( \varpi_{C_{j(k)}} \right)$. The methodology for this scale conversion is illustrated in Figure \ref{fig: scale_conversion}. The first two rows of the figure depict the comparison scale used during interviews with FEs, while the final row represents the fuzzy linguistic scale outlined in Table \ref{tab: Fuzzy scale}. For instance, when ExI is compared with another ExI criterion, the corresponding fuzzy linguistic term is ES. Similarly, if MI is compared with LI, the fuzzy linguistic equivalent is MS. This conversion process ensures that expert judgments are systematically translated into a standardized fuzzy comparison scale.

The fuzzy comparative significance $\left( \varphi_{k/(k+1)} \right)$ is calculated using Equation~\ref{eq: FUCOM3}.
\begin{align}
\varphi_{k/(k+1)} 
&= \frac{\varpi_{C_{j(k+1)}}}{\varpi_{C_{j(k)}}} 
= \frac{\left( \varpi^l_{C_{j(k+1)}},\ \varpi^m_{C_{j(k+1)}},\ \varpi^u_{C_{j(k+1)}} \right)}
         {\left( \varpi^l_{C_{j(k)}},\ \varpi^m_{C_{j(k)}},\ \varpi^u_{C_{j(k)}} \right)} \notag \\
&= \left( 
    \frac{\varpi^l_{C_{j(k+1)}}}{\varpi^u_{C_{j(k)}}},\ 
    \frac{\varpi^m_{C_{j(k+1)}}}{\varpi^m_{C_{j(k)}}},\ 
    \frac{\varpi^u_{C_{j(k+1)}}}{\varpi^l_{C_{j(k)}}} 
    \right)
= \left( \varphi^l_{k/(k+1)},\ \varphi^m_{k/(k+1)},\ \varphi^u_{k/(k+1)} \right) 
\label{eq: FUCOM3}
\end{align}

where $\varpi_{C_{j(k)}}$ represents the significance of the criterion of rank $j_k$. Hence, a fuzzy vector of comparative significance can be presented in Equation \ref{eq: FUCOM4}.
\begin{equation}
\Phi = \left( \varphi_{1/2},\ \varphi_{2/3},\ \ldots,\ \varphi_{k/(k+1)} \right)
\label{eq: FUCOM4}
\end{equation}

Lastly, the final weight coefficients are derived by solving an optimization problem while ensuring consistency through transitivity conditions. The conditions are:

\begin{itemize}
    \item Condition 1: The weight ratio coefficient between two consecutive criteria should match their comparative significance, as presented in Equation~\ref{eq: FUCOM5}.
\begin{equation}
\frac{w_k}{w_{k+1}} = \varphi_{k/(k+1)}
\label{eq: FUCOM5}
\end{equation}

where $w_k$ is the weight of the k-th criterion. 

\item Condition 2: The weight coefficients must satisfy the transitivity condition, illustrated in Equation~\ref{eq: FUCOM6}:
\begin{equation}
\frac{w_k}{w_{k+2}} = \varphi_{k/(k+1)} \otimes \varphi_{(k+1)/(k+2)}
\label{eq: FUCOM6}
\end{equation}

Thus, the final non-linear optimization model can be defined by Equation~\ref{eq: FUCOM7}.
\begin{equation}
\begin{array}{c}
\text{minimize} \quad \chi \\[1ex]
\text{s.t.} \quad 
\left\{
\begin{array}{l}
\left| \dfrac{w_k}{w_{k+1}} - \varphi_{k/(k+1)} \right| \leq \chi,\quad \forall j \\[1ex]
\left| \dfrac{w_k}{w_{k+1}} - \varphi_{k/(k+1)} \otimes \varphi_{(k+1)/(k+2)} \right| \leq \chi,\quad \forall j \\[1ex]
\displaystyle\sum_{j=1}^{n} w_j = 1,\quad \forall j \\[1ex]
w_j^l \leq w_j^m \leq w_j^u \\[1ex]
w_j^l \geq 0,\quad \forall j \\[1ex]
j = 1, 2, \ldots, n
\end{array}
\right.
\end{array}
\label{eq: FUCOM7}
\end{equation}

Where ${w}_j = (w_j^l, w_j^m, w_j^u)$ and 
$\varphi_{k/(k+1)} = (\varphi_{k/(k+1)}^l,\ \varphi_{k/(k+1)}^m,\ \varphi_{k/(k+1)}^u)$.

To achieve the highest consistency in determining the fuzzy weight coefficients, it is necessary to satisfy the following conditions, depicted in Equation~\ref{eq: FUCOM8} and Equation~\ref{eq: FUCOM9}.
\begin{align}
\frac{w_k}{w_{k+1}} - \varphi_{k/(k+1)} &= 0
\label{eq: FUCOM8} \\
\frac{w_k}{w_{k+2}} - \varphi_{k/(k+1)} \otimes \varphi_{(k+1)/(k+2)} &= 0
\label{eq: FUCOM9}
\end{align}

Therefore, the model presented in Equation~\ref{eq: FUCOM7} can be transformed into a fuzzy linear model, illustrated by Equation~\ref{eq: FUCOM10}.
\begin{equation}
\begin{array}{c}
\text{minimize} \quad \chi \\[1ex]
\text{s.t.} \quad 
\left\{
\begin{array}{l}
\left| w_k - w_{k+1} \otimes \varphi_{k/(k+1)} \right| \leq \chi, \quad \forall j \\[1ex]
\left| w_k - w_{k+1} \otimes \varphi_{k/(k+1)} \otimes \varphi_{(k+1)/(k+2)} \right| \leq \chi, \quad \forall j \\[1ex]
\sum\limits_{j=1}^{n} w_j = 1, \quad \forall j \\[1ex]
w_j^l \leq w_j^m \leq w_j^u \\[1ex]
w_j^l \geq 0, \quad \forall j \\[1ex]
j = 1, 2, \ldots, n
\end{array}
\right.
\end{array}
\label{eq: FUCOM10}
\end{equation}

By solving this linear model, the optimal fuzzy values of the weight coefficients
$\left(w_1, w_2, \ldots, w_n\right)^T$ are obtained.

\end{itemize}

\subsection{Implementation of MARCOS}
\label{Implementation of MARCOS}

The Measurement of Alternatives and Ranking according to COmpromise Solution (MARCOS) is an advanced MCDM method that evaluates alternatives based on their relationship with both an ideal and anti-ideal solution. This dual assessment enhances decision accuracy and robustness, making MARCOS a reliable approach for complex decision-making scenarios~\cite{demir2024measurement}. Unlike the Technique for Order Preference by Similarity to Ideal Solution (TOPSIS), which measures only proximity to the ideal solution, MARCOS evaluates alternatives based on their relationship with both the ideal and anti-ideal solutions, ensuring a more holistic ranking approach~\cite{chakraborty2022topsis}. Compared to VIKOR, which focuses on compromise solutions with higher sensitivity to weight variations, MARCOS provides more stable rankings by incorporating utility functions that balance both relative closeness and desirability of alternatives~\cite{opricovic2004compromise}. Although MARCOS is a new MCDM method, it is being used in diverse, complex decision-making problems, such as healthcare industries~\cite{stevic2020sustainable}, green manufacturing processes~\cite{shanmugasundar2022novel}, and mineral potential modeling~\cite{roshanravan2025bwm}. The implementation of the MARCOS method consists of seven steps. First, the decision matrix, $X$, is constructed for m alternatives and n criteria can be given by Equation~\ref{eq: MARCOS1}.

\begin{equation}
X = 
\begin{bmatrix}
x_{11} & x_{12} & \cdots & x_{1n} \\
x_{21} & x_{22} & \cdots & x_{2n} \\
\vdots & \vdots & \ddots & \vdots \\
x_{m1} & x_{m2} & \cdots & x_{mn}
\end{bmatrix}
\label{eq: MARCOS1}
\end{equation}

Second, an extended decision matrix, ${X}'$,, is formulated by including the ideal (AI) and anti-ideal (AAI) solutions and written in Equation~\ref{eq: MARCOS2}.
\begin{equation}
X' = 
\begin{bmatrix}
x_{aai,1} & x_{aai,2} & \cdots & x_{aai,n} \\
x_{11} & x_{12} & \cdots & x_{1n} \\
x_{21} & x_{22} & \cdots & x_{2n} \\
\vdots & \vdots & \ddots & \vdots \\
x_{m1} & x_{m2} & \cdots & x_{mn} \\
x_{ai,1} & x_{ai,2} & \cdots & x_{ai,n}
\end{bmatrix}
\label{eq: MARCOS2}
\end{equation}

The ideal solution (AI) corresponds to the alternative with the best performance, whereas the anti-ideal solution (AAI) represents the alternative with the worst performance. The best alternative is the one closest to the ideal solution while being farthest from the anti-ideal solution. The determination of AAI and AI depends on the type of criteria, as follows in Equation~\ref{eq: MARCOS3} and Equation~\ref{eq: MARCOS4}.
\begin{align}
\text{AAI}_j &= \min_i x_{ij}, \quad & \text{AI}_j &= \max_i x_{ij}, && \text{for benefit criteria} 
\label{eq: MARCOS3} \\
\text{AAI}_j &= \max_i x_{ij}, \quad & \text{AI}_j &= \min_i x_{ij}, && \text{for cost criteria} 
\label{eq: MARCOS4}
\end{align}

Third, the normalization of the extended matrix is performed as follows using Equation~\ref{eq: MARCOS5} and Equation~\ref{eq: MARCOS6}.
\begin{align}
n_{ij} &= \frac{x_{ai,j}}{x_{ij}}, && \text{for benefit criteria} 
\label{eq: MARCOS5} \\
n_{ij} &= \frac{x_{ij}}{x_{ai,j}}, && \text{for cost criteria} 
\label{eq: MARCOS6}
\end{align}

Fourth, the weighted normalized values in the weighted normalized matrix are computed using Equation~\ref{eq: MARCOS7}. 
\begin{equation}
v_{ij} = n_j \times {w}_j
\label{eq: MARCOS7}
\end{equation}

where $w_j$ represents the weight of the j-th criterion.

Fifth, the utility degrees for each alternative are determined with respect to AI and AAI solutions by Equation~\ref{eq: MARCOS8} and Equation~\ref{eq: MARCOS9}. 
\begin{align}
K_i^{-} &= \frac{S_i}{S_{aai}} \label{eq: MARCOS8} \\
K_i^{+} &= \frac{S_i}{S_{ai}} \label{eq: MARCOS9}
\end{align}

where $S_i$ is the sum of weighted normalized values for each alternative, which can be calculated using Equation~\ref{eq: MARCOS10}. $S_{aai}$ and $S_{ai}$  correspond to the total sum of elements in the weighted normalized matrix for the anti-ideal and ideal alternatives, respectively.
\begin{equation}
S_i = \sum_{j=1}^{n} v_{ij}
\label{eq: MARCOS10}
\end{equation}

Sixth, the utility function represents how an alternative balances between ideal and anti-ideal solutions. The utility function for each alternative is calculated by Equation~\ref{eq: MARCOS11}:
\begin{equation}
f(K_i) = \frac{K_i^+ + K_i^-}{1 + \dfrac{1 - f(K_i^+)}{f(K_i^+)} + \dfrac{1 - f(K_i^-)}{f(K_i^-)}}
\label{eq: MARCOS11}
\end{equation}

where, 
\begin{align}
f(K_i^-) &= \frac{K_i^+}{K_i^+ + K_i^-} \label{eq:f_Kminus} \\
f(K_i^+) &= \frac{K_i^-}{K_i^+ + K_i^-} \label{eq:f_Kplus}
\end{align}

Finally, alternatives are ranked based on their final utility function values, with the highest value indicating the best alternative.

\subsection{Sensitivity analysis}

The following steps outline the approach used in this research to determine the most critical decision criteria. First, the criteria weights and the decision matrix are obtained from the MARCOS implementation. The normalized decision matrix is formulated by removing the AI and AAI solutions from the extended normalized matrix. Then, the preference matrix, $f(K_i)$, is also obtained from the MARCOS implementation.

After that, to evaluate the sensitivity of each criterion, we compute the minimum required change in the weight of a criterion $w_k$ that would result in a reversal of rankings between alternatives $A_i$ and $A_j$. This is represented by $\delta_{kij}$, which is derived as follows in Equation~\ref{eq: sensitivity1} and Equation~\ref{eq: sensitivity2}~\cite{triantaphyllou2000multi}. 
\begin{equation}
\delta_{kij} < \left( \frac{P_j - P_i}{a_{jk} - a_{ik}} \right) \times \frac{100}{w_k}, \quad \text{if } (a_{jk} > a_{ik})
\label{eq: sensitivity1}
\end{equation}

or,

\begin{equation}
\delta_{kij} > \left( \frac{P_j - P_i}{a_{jk} - a_{ik}} \right) \times \frac{100}{w_k}, \quad \text{if } (a_{jk} < a_{ik})
\label{eq: sensitivity2}
\end{equation}

where $P_i$ and $P_j$ denote the preference values of alternatives $A_i$ and $A_j$, respectively. The feasibility condition of the above equations is defined in Equation~\ref{eq: sensitivity3}.

\begin{equation}
\frac{P_j - P_i}{a_{jk} - a_{ik}} \leq w_k
\label{eq: sensitivity3}
\end{equation}

Given 85 alternatives (coal sites), a total of 3,570 pairwise comparisons are conducted to compute the values of $\delta_{kij}$.

Lastly, the criticality degree of each criterion $D_k'$ is computed by Equation~\ref{eq: sensitivity4}~\cite{triantaphyllou2000multi}:
\begin{equation}
D_k' = \max_{1 \leq i \leq j \leq m} \left( |\delta_{kij}| \right), \quad \forall\ 1 \leq k \leq n
\label{eq: sensitivity4}
\end{equation}

This value represents the minimum percentage change required in $w_k$ to alter the ranking of alternatives.

\section{RESULTS}


\subsection{Determination of weights of the attribute and sub-attribute for fusion siting}



Based on the previously defined equations (see Methods section and Supplementary Information) and expert evaluations, we formulated an optimization problem to determine the weight coefficients of the main criteria: State Policies (SP), Federal Policies (FP), Risk \& Hazard Metrics (RHM), and Connectivity \& Spatial Factors (CSF). Since experts did not directly rank the main criteria in their decision-making process, we computed the significance value of each main criterion by summing the scale points of its sub-criteria and taking the average. The ranking and comparative significance of the criteria are presented in Table \ref{tab:fe_ranking_phi}. Table~\ref{tab:fe_ranking_phi} also includes the comparative priority vector ($\Phi$). The details of all the mathematical notations used in this section are explained in the Methods section.

\begin{table}[htbp]
\centering
\small
\caption{Details of individual Fusion Experts (FEs) ranking, comparative significance of the criteria imposed by the FEs, and comparative priority vector of different attributes.}
\label{tab:fe_ranking_phi}
\small
\begin{tabular}{|M{1cm}|M{5cm}|M{4.1cm}|M{4.4cm}|}
\hline
\textbf{FEs} & \textbf{Individual expert ranking} & \textbf{Comparative significance} ($\bm{\varpi_{C_{j(k)}}}$) & \textbf{Comparative priority vector} ($\bm{\Phi}$) \\
\hline
FE1 & FP (3.67) $>$ RHM (3.56) $>$ CSF (3.40) $>$ SP (3.00) & \{1.00, 1.03, 1.08, 1.22\} & \{1.03, 1.05, 1.13\} \\
\hline
FE2 & CSF (3.80) $>$ SP (3.40) $>$ FP (3.30) $>$ RHM (2.00) & \{1.00, 1.12, 1.14, 1.90\} & \{1.12, 1.02, 1.67\} \\
\hline
FE3 & RHM (3.33) $>$ FP (3.00) = CSF (3.00) $>$ SP (2.00) & \{1.00, 1.11, 1.11, 1.67\} & \{1.11, 1.00, 1.50\} \\
\hline
FE4 & RHM (3.56) $>$ FP (3.00) $>$ CSF (2.80) $>$ SP (2.00) & \{1.00, 1.19, 1.27, 1.78\} & \{1.19, 1.07, 1.40\} \\
\hline
FE5 & CSF (4.00) $>$ RHM (3.44) $>$ SP (3.20) $>$ FP (3.00) & \{1.00, 1.16, 1.25, 1.33\} & \{1.16, 1.08, 1.06\} \\
\hline
\end{tabular}

\end{table}

Since we calculated specific numerical values for each main criterion without linguistic vagueness, we employed the FUCOM method rather than F-FUCOM in this case. Since responses were collected from five fusion experts, we formulated five separate optimization problems, each corresponding to an expert's input. The detailed formulations are presented in the supplemental information (see Note S2). The optimization problems in this research were solved using the “gurobi” solver. After solving each problem, we obtained the weights of each attribute as presented in Table~\ref{tab: optimal_weights}. In all cases, we also achieved maximum consistency ($\chi$=0).

\begin{table}[htbp]
\centering
\small
\caption{Optimal weights of different attributes.}
\begin{tabular}{|c|c|c|}
\hline
\textbf{FEs} & \textbf{Weights \{{SP}, {FP}, {RHM}, {CSF}\}} & $\bm{\chi}$ \\
\hline
FE1 & \{0.2206, 0.2691, 0.2612, 0.2491\} & 0 \\
\hline
FE2 & \{0.2709, 0.2661, 0.1597, 0.3034\} & 0 \\
\hline
FE3 & \{0.1761, 0.2649, 0.2941, 0.2649\} & 0 \\
\hline
FE4 & \{0.1761, 0.2635, 0.3135, 0.2469\} & 0 \\
\hline
FE5 & \{0.2343, 0.2202, 0.2525, 0.2929\} & 0 \\
\hline
\end{tabular}
\label{tab: optimal_weights}
\end{table}

Next, the F-FUCOM method was used to determine the weights of each sub-criterion of SP. Table \ref{tab: Details of FEs’ input on different SP sub-attributes} presents the input provided by the fusion experts, including their linguistic evaluations and the corresponding numerical scale values (shown in parentheses). Based on the implementation steps described in the Methods section, the resulting comparative ranking and significance imposed by each fusion expert are summarized in Table~\ref{tab: Details of individual comparative ranking on SP sub-attributes} and Table~\ref{tab:FE on SP sub-attributes}.

\begin{table}[htbp]
\centering
\small
\caption{Details of FEs’ input on different SP sub-attributes.}
\label{tab: Details of FEs’ input on different SP sub-attributes}

\begin{tabular}{|M{0.8cm}|M{2.2cm}|M{2.5cm}|M{1.8cm}|M{2.2cm}|M{2.4cm}|M{2.2cm}|}

\hline
 \textbf{FEs} &  \textbf{Nuclear  Restriction (SP1)} &  \textbf{Nuclear Inclusive Policy (SP2)} &  \textbf{Energy Price (SP3)} &  \textbf{Market Regulation (SP4)} &  \textbf{Construction Labor Rate (SP5)}\\ 
 \hline
 FE1 &  LI (2)&  MI (3)&  I (4)&  MI (3)&  MI (3)\\ 
 \hline
 FE2 &  MI (3)&  ExI (5)&  ExI (5)&  LI (2)&  LI (2)\\ 
 \hline
 FE3
&  NI (1)&  NI (1)&  I (4)&  LI (2)&  LI (2)\\ 
\hline
 FE4
&  NI (1)&  LI (2)&  LI (2)&  LI (2)&  MI (3)\\ 
\hline
 FE5
&  LI (2)&  MI (3)&  ExI (5)&  MI (3)&  MI (3)\\ 
\hline
\end{tabular}
\label{tab:blank_7x6}
\end{table}

\begin{table}[htbp]
\centering
\small
\caption{Details of individual comparative ranking on SP sub-attributes.}
\label{tab: Details of individual comparative ranking on SP sub-attributes}

\begin{tabular}{|c|c|}
\hline
\textbf{FEs} & \textbf{Ranking by the individual expert} \\
\hline
FE1 & SP3 $>$ SP2 = SP4 = SP5 $>$ SP1 \\
\hline
FE2 & SP2 = SP3 $>$ SP1 $>$ SP4 = SP5 \\
\hline
FE3 & SP3 $>$ SP4 = SP5 $>$ SP1 = SP2 \\
\hline
FE4 & SP5 $>$ SP2 = SP3 = SP4 $>$ SP1 \\
\hline
FE5 & SP3 $>$ SP2 = SP4 = SP5 $>$ SP1 \\
\hline
\end{tabular}
\end{table}

\begin{table}[htbp]
\centering
\small
\caption{Details of comparative significance imposed by individual FE on SP sub-attributes}
\label{tab:FE on SP sub-attributes}

\begin{tabular}{|M{0.8cm}|M{5.4cm}|M{8.1cm}|}
\hline
\textbf{FEs} & 
\textbf{Significance (linguistic) of the criteria imposed by the individual expert ($\bm{\varpi_{C_{j(k)}}}$)} & 
\textbf{Significance (TFN) of the criteria imposed by the individual expert ($\bm{\varpi_{C_{j(k)}}}$)} \\
\hline
FE1 & \{ES, MS, MS, MS, SS\} & \{(1,1,1), (2,3,4), (2,3,4), (2,3,4), (4,5,6)\} \\
\hline
FE2 & \{ES, ES, SS, VSS, VSS\} & \{(1,1,1), (1,1,1), (4,5,6), (6,7,8), (6,7,8)\} \\
\hline
FE3 & \{ES, SS, SS, VSS, VSS\} & \{(1,1,1), (4,5,6), (4,5,6), (6,7,8), (6,7,8)\} \\
\hline
FE4 & \{ES, MS, MS, MS, SS\} & \{(1,1,1), (2,3,4), (2,3,4), (2,3,4), (4,5,6)\} \\
\hline
FE5 & \{ES, SS, SS, SS, VSS\} & \{(1,1,1), (4,5,6), (4,5,6), (4,5,6), (6,7,8)\} \\
\hline
\end{tabular}
\end{table}

Then, we formulated a distinct optimization problem for each fusion expert based on their input in Table~\ref{tab: Details of FEs’ input on different SP sub-attributes}, aiming to minimize the consistency deviation parameter ($\chi$) and derive the optimal weights for the sub-criteria. The complete formulation of the optimization model is presented in the Supplement Information (see Note S3). After solving the optimization problems, the resulting fuzzy weights for each sub-criterion are presented in Table~\ref{tab:Optimal fuzzy weights of different SP sub-attributes}.

\begin{table}[htbp]
\centering
\caption{Optimal fuzzy weights of different SP sub-attributes}
\label{tab:Optimal fuzzy weights of different SP sub-attributes}
\small
\renewcommand{\arraystretch}{1.4} 
\begin{tabular}{|M{0.8cm}|M{2.3cm}|M{2.3cm}|M{2.3cm}|M{2.3cm}|M{2.3cm}|M{2.3cm}|}
\hline
\multirow{2}{*}{\textbf{FEs}} 
& \multicolumn{5}{c|}{\textbf{Fuzzy weights}} \\ 
\cline{2-6}
& \textbf{SP1} & \textbf{SP2} & \textbf{SP3} & \textbf{SP4} & \textbf{SP5}\\ 
\hline
FE1 & (0.038, 0.115, 0.115) & (0.107, 0.147, 0.147) & (0.231, 0.493, 0.493) & (0.069, 0.168, 0.168) & (0.052, 0.178, 0.178) \\
\hline
FE2 & (0.068,    0.082,    0.096) & (0.385,    0.385,    0.385) & (0.361,    0.409,    0.409) & (0.036,    0.062,    0.084) & (0.045,    0.076,    0.080)\\
\hline
FE3 & (0.049,    0.104,    0.118) & (0.065,    0.115,    0.115) & (0.397,    0.581,    0.690) & (0.109,    0.109,    0.109) & (0.081,    0.124,    0.135)\\
\hline
FE4 & (0.038,    0.115,    0.115) & (0.107,    0.147,    0.147) & (0.069,    0.168,    0.168) & (0.052,    0.178,    0.178) & (0.231    0.493    0.493)\\
\hline
FE5 & (0.056,    0.100,    0.100) & (0.103,    0.107,    0.107) & (0.390,    0.570,    0.656) & (0.077,    0.121,    0.132) & (0.064,    0.143,    0.149)\\
\hline
\end{tabular}
\end{table}

To derive the final crisp weights from Table~\ref{tab:Optimal fuzzy weights of different SP sub-attributes}, we applied the Graded Mean Integration Representation (GMIR) method, as shown in Equation~\ref{eq:GMIR} \cite{haqbin2022comparing}.

\begin{equation}
C(x_j) = \frac{l_j + 4m_j + u_j}{6}
\label{eq:GMIR}
\end{equation}

Where $x_j = (l_j, m_j, u_j)$ is a triangular fuzzy number.

Table~\ref{tab: Optimal crisp weights of different SP sub-attributes} shows that we obtained the consistency deviation $\chi$ less than 0.10 in all cases, implying an acceptable deviation range from full consistency. The computed crisp weights are summarized in Table~\ref{tab: Optimal crisp weights of different SP sub-attributes}.

\begin{table}[htbp]
\centering
\caption{Optimal crisp weights of different SP sub-attributes}
\label{tab: Optimal crisp weights of different SP sub-attributes}
\small
\begin{tabular}{|c|c|c|c|c|c|c|}
\hline
\multirow{2}{*}{\textbf{FEs}} & \multicolumn{5}{c|}{\textbf{Crisp weights}} & \multirow{2}{*}{$\chi$} \\
\cline{2-6}
& \textbf{SP1} & \textbf{SP2} & \textbf{SP3} & \textbf{SP4} & \textbf{SP5} & \\
\hline
FE1 & 0.103 & 0.140 & 0.449 & 0.151 & 0.157 & 0.06 \\
\hline
FE2 & 0.082 &	0.385 &	0.401 &	0.061 &	0.071 & 0.02 \\
\hline
FE3 & 0.098 &	0.107 &	0.569 &	0.109 &	0.119 & 0.04 \\
\hline
FE4 & 0.103 & 0.140 &	0.151 &	0.157 &	0.449 & 0.06 \\
\hline
FE5 & 0.093 &	0.106 &	0.554 &	0.116 &	0.131 & 0.04 \\
\hline
\end{tabular}
\end{table}

Similarly, we determined the weights of the sub-criteria under the FP category. The experts’ evaluations are provided in Table~\ref{tab: Details of FEs’ input on different FP sub-attributes}, with their comparative rankings and significance levels presented in Table~\ref{tab: Details of individual comparative ranking on FP sub-attributes} and Table~\ref{tab:Details of comparative significance imposed by individual FE on FP sub-attributes}, respectively. We formulated optimization problems, detailed in the Supplement Information (see Note S4), based on these inputs to derive the optimal weights of FP's sub-criteria while minimizing inconsistency. The resulting fuzzy and corresponding crisp weights are reported in Table~\ref{tab: Optimal fuzzy weights of different FP sub-attributes} and Table~\ref{tab: Optimal crisp weights of different FP sub-attributes}, respectively. Table~\ref{tab: Optimal crisp weights of different FP sub-attributes} shows that optimization achieves full consistency, with all $\chi$ values equal to zero.

\begin{table}[htbp]
\centering

\caption{Details of FEs’ input on different FP sub-attributes}
\label{tab: Details of FEs’ input on different FP sub-attributes}
\small
\begin{tabular}{|M{0.8cm}|M{3cm}|M{3cm}|M{3.4cm}|}

\hline
 \textbf{FEs} &  \textbf{Net Electricity Imports (FP1)} &  \textbf{Hydrogen Demand (FP2)} &  \textbf{Federal Incentives (FP3)}\\ 
 \hline
 FE1 &  I (4)&  MI (3)&  I (4)\\ 
 \hline
 FE2 &  I (4)& NI (1)&  ExI (5)\\ \hline
 FE3 &  MI (3)&  NI (1)&  ExI (5)\\ \hline
 FE4 &  MI (3)&  NI (1)&  ExI (5)\\ \hline
 FE5 &  MI (3)&  LI (2)&  I (4)\\ \hline
\end{tabular}
\end{table}

\begin{table}[htbp]
\centering
\small
\caption{Details of individual comparative ranking on FP sub-attributes}
\label{tab: Details of individual comparative ranking on FP sub-attributes}

\begin{tabular}{|c|c|}
\hline
\textbf{FEs} & \textbf{Ranking by the individual expert} \\
\hline
FE1 & FP1 = FP3 $>$ FP2 \\
\hline
FE2 & FP3 $>$ FP1 $>$ FP2   \\
\hline
FE3 & FP3 $>$ FP1 $>$ FP2 \\
\hline
FE4 & FP3 $>$ FP1 $>$ FP2 \\
\hline
FE5 & FP3 $>$ FP1 $>$ FP2 \\
\hline
\end{tabular}
\end{table}

\begin{table}[htbp]
\centering
\small
\caption{Details of comparative significance imposed by individual FE on FP sub-attributes}
\label{tab:Details of comparative significance imposed by individual FE on FP sub-attributes}

\begin{tabular}{|M{0.8cm}|M{5.4cm}|M{8.1cm}|}
\hline
\textbf{FEs} & 
\textbf{Significance (linguistic) of the criteria imposed by the individual expert ($\bm{\varpi_{C_{j(k)}}}$)} & 
\textbf{Significance (TFN) of the criteria imposed by the individual expert ($\bm{\varpi_{C_{j(k)}}}$)} \\
\hline
FE1 & \{ES,ES,MS\} & \{(1,1,1),(1,1,1),(2,3,4)\} \\
\hline
FE2 & \{ES,MS,ExS\} & \{(1,1,1),(2,3,4),(9,9,9)\} \\
\hline
FE3 & \{ES,SS,ExS\} & \{(1,1,1),(4,5,6),(9,9,9)\} \\
\hline
FE4 & \{ES,SS,ExS\} & \{(1,1,1),(4,5,6),(9,9,9)\} \\
\hline
FE5 & \{ES,MS,SS\} & \{(1,1,1),(2,3,4),(4,5,6)\} \\
\hline
\end{tabular}
\end{table}

\begin{table}[htbp]
\centering
\caption{Optimal fuzzy weights of different FP sub-attributes}
\label{tab: Optimal fuzzy weights of different FP sub-attributes}

\small

\renewcommand{\arraystretch}{1.4} 
\begin{tabular}{|M{0.8cm}|M{3.8cm}|M{3.8cm}|M{3.8cm}|}
\hline
\multirow{2}{*}{\textbf{FEs}} 
& \multicolumn{3}{c|}{\textbf{Fuzzy weights}} \\ 
\cline{2-4}
& \textbf{FP1} & \textbf{FP2} & \textbf{FP3} \\ 
\hline
FE1 & (0.426, 0.426, 0.426) & (0.107, 0.142, 0.213) & (0.426, 0.426, 0.426) \\
\hline
FE2 & (0.231, 0.231, 0.231) & (0.051, 0.077, 0.103) & (0.462, 0.692, 0.923) \\
\hline
FE3 & (0.153, 0.153, 0.153) & (0.068, 0.085, 0.102) & (0.610, 0.763, 0.915) \\
\hline
FE4 & (0.153, 0.153, 0.153) & (0.068, 0.085, 0.102) & (0.610, 0.763, 0.915) \\
\hline
FE5 & (0.216, 0.216, 0.216) & (0.072, 0.130, 0.216) & (0.433, 0.649, 0.865)  \\
\hline
\end{tabular}
\end{table}

\begin{table}[htbp]
\centering
\caption{Optimal crisp weights of different FP sub-attributes}
\label{tab: Optimal crisp weights of different FP sub-attributes}
\small
\begin{tabular}{|c|c|c|c|c|}
\hline
\multirow{2}{*}{\textbf{FEs}} & \multicolumn{3}{c|}{\textbf{Crisp weights}} & \multirow{2}{*}{$\chi$} \\
\cline{2-4}
& \textbf{FP1} & \textbf{FP2} & \textbf{FP3} & \\
\hline
FE1 & 0.426 & 0.148 & 0.426 & 0.00 \\
\hline
FE2 & 0.231 & 0.077 & 0.692 & 0.00 \\
\hline
FE3 & 0.153 & 0.085 & 0.763 & 0.00 \\
\hline
FE4 & 0.153 & 0.085 & 0.763 & 0.00 \\
\hline
FE5 & 0.216 & 0.135 & 0.649 & 0.00 \\
\hline
\end{tabular}
\end{table}

Next, we determined the optimal weights of the sub-criteria for the RHM category. The experts’ inputs are presented in Table~\ref{tab: Details of FEs’ input on different RHM sub-attributes}, while their comparative rankings and importance levels are shown in Table~\ref{tab: Details of individual comparative ranking on RHM sub-attributes} and Table~\ref{tab:Details of comparative significance imposed by individual FE on RHM sub-attributes}, respectively. We formulated and solved optimization problems, as detailed in the Supplement Information (see Note S5), using these inputs to obtain optimal weights with minimal inconsistency. The resulting fuzzy weights and their corresponding crisp values are provided in Table~\ref{tab: Optimal fuzzy weights of different RHM sub-attributes} and Table~\ref{tab: Optimal crisp weights of different RHM sub-attributes}, respectively. As indicated in Table~\ref{tab: Optimal crisp weights of different RHM sub-attributes}, all $\chi$ values remain below 0.10, reflecting an acceptable level of consistency among the RHM sub-criteria.

\begin{table}[htbp]
\centering
\caption{Details of FEs’ input on different RHM sub-attributes}
\label{tab: Details of FEs’ input on different RHM sub-attributes}
\small
\begin{tabular}{|M{0.8cm}|M{1.7cm}|M{1.9cm}|M{1.3cm}|M{1.7cm}|M{1.8cm}|M{1.3cm}|M{1.8cm}|M{1.3cm}|}
\hline
 \textbf{FEs} &  \textbf{Protected Lands (RHM1)} &  \textbf{Hazardous Facilities (RHM2)} &  \textbf{Fault Lines (RHM3)} &  \textbf{Landslide Hazard (RHM4)} &  \textbf{Safe Shutdown Earthquake (RHM5)} &  \textbf{100-Years Flood (RHM6)} & \textbf{Open Water \& Wetlands (RHM7)} & \textbf{Slope (RHM8)}\\ 
 \hline
 FE1 &  MI (3) & MI (3) & I (4) & I (4) & MI (3) & I (4) & I (4) & MI (3)\\ 
 \hline
 FE2 &  NI (1) & NI (1) & MI (3) & MI (3) & NI (1) & MI (3) & NI (1) & LI (2)
\\ \hline
 FE3 &  MI (3) & I (4) & MI (3) & I (4) & MI (3) & MI (3) & I (4) & MI (3)
\\ \hline
 FE4
&  MI (3) & LI (2) & I (4) & I (4) & I (4) & I (4) & MI (3) & I (4)
\\ \hline
 FE5
&  MI (3) & I (4) & MI (3) & I (4) & I (4) & I (4) & MI (3) & MI (3)
\\ \hline
\end{tabular}
\end{table}

\begin{table}[htbp]
\centering
\small
\caption{Details of individual comparative ranking on RHM sub-attributes}
\label{tab: Details of individual comparative ranking on RHM sub-attributes}

\begin{tabular}{|c|c|}
\hline
\textbf{FEs} & \textbf{Ranking by the individual expert} \\
\hline
FE1 & RHM3 = RHM4 = RHM6 = RHM7 $>$ RHM1 = RHM2 = RHM5 = RHM8 \\
\hline
FE2 & RHM3 = RHM4 = RHM6 $>$ RHM8 $>$ RHM1 = RHM2 = RHM5 = RHM7   \\
\hline
FE3 & RHM2 = RHM4 = RHM7 $>$ RHM1 = RHM3 = RHM5 = RHM6 = RHM8 \\
\hline
FE4 & RHM3 = RHM4 = RHM5 = RHM6 = RHM8 $>$ RHM1 = RHM7 $>$ RHM2 \\
\hline
FE5 & RHM2 = RHM4 = RHM5 = RHM6 $>$ RHM1 = RHM3 = RHM7 = RHM8 \\
\hline
\end{tabular}
\end{table}

\begin{table}[htbp]
\centering
\small
\caption{Details of comparative significance imposed by individual FE on RHM sub-attributes}
\label{tab:Details of comparative significance imposed by individual FE on RHM sub-attributes}

\begin{tabular}{|M{0.8cm}|M{5.4cm}|M{9.8cm}|}
\hline
\textbf{FEs} & 
\textbf{Significance (linguistic) of the criteria imposed by the individual expert ($\bm{\varpi_{C_{j(k)}}}$)} & 
\textbf{Significance (TFN) of the criteria imposed by the individual expert ($\bm{\varpi_{C_{j(k)}}}$)} \\
\hline
FE1 & \{ES,ES,ES,ES,MS,MS,MS,MS\} & \{(1,1,1),(1,1,1),(1,1,1),(1,1,1),(2,3,4),(2,3,4),(2,3,4),(2,3,4)\} \\
\hline
FE2 & \{ES,ES,ES,MS,SS,SS,SS,SS\} & \{(1,1,1),(1,1,1),(1,1,1),(2,3,4),(4,5,6),(4,5,6),(4,5,6),(4,5,6)\} \\
\hline
FE3 & \{ES,ES,ES,MS,MS,MS,MS,MS\} & \{(1,1,1),(1,1,1),(1,1,1),(2,3,4),(2,3,4),(2,3,4),(2,3,4),(2,3,4)\} \\
\hline
FE4 & \{ES,ES,ES,ES,ES,MS,MS,SS\} & \{(1,1,1),(1,1,1),(1,1,1),(1,1,1),(1,1,1),(2,3,4),(2,3,4),(4,5,6)\} \\
\hline
FE5 & \{ES,ES,ES,ES,MS,MS,MS,MS\} & \{(1,1,1),(1,1,1),(1,1,1),(1,1,1),(2,3,4),(2,3,4),(2,3,4),(2,3,4)\} \\
\hline
\end{tabular}
\end{table}

\begin{table}[htbp]
\centering
\caption{Optimal fuzzy weights of different RHM sub-attributes}
\label{tab: Optimal fuzzy weights of different RHM sub-attributes}

\small

\renewcommand{\arraystretch}{1.4} 
\begin{tabular}{|M{0.8cm}|M{1.4cm}|M{1.4cm}|M{1.4cm}|M{1.4cm}|M{1.4cm}|M{1.4cm}|M{1.4cm}|M{1.4cm}|}
\hline
\multirow{2}{*}{\textbf{FEs}} 
& \multicolumn{8}{c|}{\textbf{Fuzzy weights}} \\ 
\cline{2-9}
& \textbf{RHM1} & \textbf{RHM2} & \textbf{RHM3} & \textbf{RHM4} & \textbf{RHM5} & \textbf{RHM6} & \textbf{RHM7} & \textbf{RHM8}\\ 
\hline
FE1 & (0.048, 0.064, 0.079) & (0.028, 0.075, 0.090) & (0.191, 0.191, 0.191) & (0.157, 0.157, 0.157) & (0.028, 0.095, 0.095) & (0.157, 0.157, 0.157) & (0.123, 0.191, 0.191) & (0.031, 0.109, 0.123) \\
\hline
FE2 & (0.022, 0.053, 0.073) & (0.028, 0.066, 0.079) & (0.206, 0.206, 0.206) & (0.206, 0.206, 0.206) & (0.046, 0.084, 0.088) & (0.176, 0.237, 0.237) & (0.049, 0.096, 0.114) & (0.059, 0.079, 0.097) \\
\hline
FE3 & (0.052, 0.070, 0.086) & (0.172, 0.172, 0.172) & (0.030, 0.082, 0.099) & (0.172, 0.172, 0.172) & (0.030, 0.105, 0.105) & (0.034, 0.108, 0.108) & (0.136, 0.209, 0.209) & (0.035, 0.142, 0.142) \\
\hline
FE4 & (0.045, 0.060, 0.074) & (0.018, 0.055, 0.058) & (0.180, 0.180, 0.180) & (0.180, 0.180, 0.180) & (0.148, 0.148, 0.148) & (0.148, 0.148, 0.148) & (0.027, 0.071, 0.085) & (0.116, 0.180, 0.180) \\
\hline
FE5 & (0.048, 0.064, 0.079) & (0.191, 0.191, 0.191) & (0.028, 0.075, 0.090) & (0.157, 0.157, 0.157) & (0.157, 0.157, 0.157) & (0.123, 0.191, 0.191) & (0.028, 0.095, 0.095) & (0.031, 0.109, 0.123) \\
\hline
\end{tabular}
\end{table}

\begin{table}[htbp]
\centering
\caption{Optimal crisp weights of different RHM sub-attributes}
\label{tab: Optimal crisp weights of different RHM sub-attributes}
\small
\begin{tabular}{|c|c|c|c|c|c|c|c|c|c|}

\hline
\multirow{2}{*}{\textbf{FEs}} & \multicolumn{8}{c|}{\textbf{Crisp weights}} & \multirow{2}{*}{$\chi$} \\
\cline{2-9}
& \textbf{RHM1} & \textbf{RHM2} & \textbf{RHM3} & \textbf{RHM4} & \textbf{RHM5} & \textbf{RHM6} & \textbf{RHM7} & \textbf{RHM8} & \\
\hline
FE1 & 0.064 & 0.070 & 0.191 & 0.157 & 0.084 & 0.157 & 0.180 & 0.098 & 0.03 \\
\hline
FE2 & 0.051 & 0.062 & 0.206 & 0.206 & 0.078 & 0.227 & 0.091 & 0.079 & 0.03 \\
\hline
FE3 & 0.070 & 0.173 & 0.076 & 0.173 & 0.092 & 0.096 & 0.197 & 0.124 & 0.04 \\
\hline
FE4 & 0.060 & 0.049 & 0.180 & 0.180 & 0.148 & 0.148 & 0.066 & 0.169 & 0.03 \\
\hline
FE5 & 0.064 & 0.191 & 0.070 & 0.157 & 0.157 & 0.180 & 0.084 & 0.098 & 0.03 \\
\hline
\end{tabular}
\end{table}

Finally, we determined the weights of the sub-criteria under the CSF category. The experts’ inputs are outlined in Table~\ref{tab: Details of FEs’ input on different CSF sub-attributes}, while their comparative rankings and significance levels are shown in Table~\ref{tab: Details of individual comparative ranking on CSF sub-attributes} and Table~\ref{tab:Details of comparative significance imposed by individual FE on CSF sub-attributes}, respectively. We formulated optimization problems, presented in the supplement information (see Note S6), using this information to derive optimal weights of CSF's sub-criteria with minimal inconsistency. Table~\ref{tab: Optimal fuzzy weights of different CFS sub-attributes} and Table~\ref{tab: Optimal crisps weights of different CSF sub-attributes} present the resulting fuzzy weights and the corresponding crisp weights, respectively. Table~\ref{tab: Optimal crisps weights of different CSF sub-attributes} indicates that the solution yields an acceptable deviation from full consistency, with all $\chi$ values remaining below 0.10.

\begin{table}[htbp]
\centering

\caption{Details of FEs’ input on different CSF sub-attributes}
\label{tab: Details of FEs’ input on different CSF sub-attributes}
\small
\begin{tabular}{|M{0.8cm}|M{2cm}|M{2.8cm}|M{2.5cm}|M{1.7cm}|M{2cm}|}

\hline
 \textbf{FEs} &  \textbf{Population (CSF1)} &  \textbf{Transportation (CSF2)} &  \textbf{Operating Nuclear facilities (CSF3)} &  \textbf{Nuclear R\&D (CSF4)} &  \textbf{Substation
(CSF5)}\\ 
 \hline
 FE1 &  I (4) & I (4) & MI (3) & LI (2) & I (4)\\ 
 \hline
 FE2 &  MI (3) & I (4) & I (4) & MI (3) & ExI (5)
\\ \hline
 FE3 &  LI (2) & ExI (5) & LI (2) & LI (2) & I (4)
\\ \hline
 FE4 &  LI (2) & ExI (5) & LI (2) & LI (2) & MI (3)
\\ \hline
 FE5
&  MI (3) & I (4) & I (4) & I (4) & ExI (5)
\\ \hline
\end{tabular}
\end{table}

\begin{table}[htbp]
\centering
\small
\caption{Details of individual comparative ranking on CSF sub-attributes}
\label{tab: Details of individual comparative ranking on CSF sub-attributes}

\begin{tabular}{|c|c|}
\hline
\textbf{FEs} & \textbf{Ranking by the individual expert} \\
\hline
FE1 & CSF1 = CSF2 = CSF5 $>$ CSF3 $>$ CSF4 \\
\hline
FE2 & CSF5 $>$ CSF2 = CSF3 $>$ CSF1 = CSF4   \\
\hline
FE3 & CSF2 $>$ CSF5 $>$ CSF1 = CSF3 = CSF4 \\
\hline
FE4 &CSF2 $>$ CSF5 $>$ CSF1 = CSF3 = CSF4 \\
\hline
FE5 & CSF5 $>$ CSF2 = CSF3 = CSF4 $>$ CSF1 \\
\hline
\end{tabular}
\end{table}

\begin{table}[htbp]
\centering
\small
\caption{Details of comparative significance imposed by individual FE on CSF sub-attributes}
\label{tab:Details of comparative significance imposed by individual FE on CSF sub-attributes}

\begin{tabular}{|M{0.8cm}|M{5.4cm}|M{9.8cm}|}
\hline
\textbf{FEs} & 
\textbf{Significance (linguistic) of the criteria imposed by the individual expert ($\bm{\varpi_{C_{j(k)}}}$)} & 
\textbf{Significance (TFN) of the criteria imposed by the individual expert ($\bm{\varpi_{C_{j(k)}}}$)} \\
\hline
FE1 & \{ES,ES,ES,MS,SS\} & \{(1,1,1),(1,1,1),(1,1,1),(2,3,4),(4,5,6)\} \\
\hline
FE2 & \{ES,MS,MS,SS,SS\} & \{(1,1,1),(2,3,4),(2,3,4),(4,5,6),(4,5,6)\} \\
\hline
FE3 & \{ES,MS,VSS,VSS,VSS\} & \{(1,1,1),(2,3,4),(6,7,8),(6,7,8),(6,7,8)\} \\
\hline
FE4 & \{ES,SS,VSS,VSS,VSS\} & \{(1,1,1),(4,5,6),(6,7,8),(6,7,8),(6,7,8)\} \\
\hline
FE5 & \{ES,MS,MS,MS,SS\} & \{(1,1,1),(2,3,4),(2,3,4),(2,3,4),(4,5,6)\} \\
\hline
\end{tabular}
\end{table}

\begin{table}[htbp]
\centering
\caption{Optimal fuzzy weights of different CFS sub-attributes}
\label{tab: Optimal fuzzy weights of different CFS sub-attributes}

\small

\renewcommand{\arraystretch}{1.4} 
\begin{tabular}{|M{0.8cm}|M{2.4cm}|M{2.4cm}|M{2.4cm}|M{2.4cm}|M{2.4cm}|}
\hline
\multirow{2}{*}{\textbf{FEs}} 
& \multicolumn{5}{c|}{\textbf{Fuzzy weights}} \\ 
\cline{2-6}
& \textbf{CFS1} & \textbf{CFS2} & \textbf{CFS3} & \textbf{CFS4} & \textbf{CFS5} \\ 
\hline
FE1 & (0.282, 0.282, 0.282) & (0.250, 0.282, 0.282) & (0.078, 0.104, 0.110) & (0.026, 0.063, 0.110) & (0.250, 0.282, 0.282) \\
\hline
FE2 & (0.035, 0.125, 0.139) & (0.132, 0.145, 0.145) & (0.081, 0.166, 0.166) & (0.051, 0.137, 0.145) & (0.229, 0.499, 0.589) \\
\hline
FE3 & (0.047, 0.088, 0.093) & (0.318, 0.574, 0.706) & (0.041, 0.093, 0.114) & (0.074, 0.106, 0.106) & (0.167, 0.178, 0.178) \\
\hline
FE4 & (0.062, 0.091, 0.101) & (0.432, 0.605, 0.720) & (0.054, 0.103, 0.122) & (0.073, 0.110, 0.110) & (0.115, 0.115, 0.115) \\
\hline
FE5 & (0.038, 0.115, 0.115) & (0.107, 0.147, 0.147) & (0.069, 0.168, 0.168) & (0.052, 0.178, 0.178) & (0.231, 0.493, 0.493) \\
\hline
\end{tabular}
\end{table}

\begin{table}[htbp]
\centering
\small
\caption{Optimal crisps weights of different CSF sub-attributes}
\label{tab: Optimal crisps weights of different CSF sub-attributes}
\begin{tabular}{|c|c|c|c|c|c|c|}
\hline
\multirow{2}{*}{\textbf{FEs}} & \multicolumn{5}{c|}{\textbf{Crisp weights}} & \multirow{2}{*}{$\chi$} \\
\cline{2-6}
& \textbf{CSF1} & \textbf{CSF2} & \textbf{CSF3} & \textbf{CSF4} & \textbf{CSF5} & \\
\hline
FE1 & 0.282 & 0.277 & 0.101 & 0.064 & 0.277 & 0.03 \\
\hline
FE2 & 0.112 & 0.143 & 0.152 & 0.124 & 0.469 & 0.06 \\
\hline
FE3 & 0.082 & 0.553 & 0.088 & 0.101 & 0.177 & 0.04 \\
\hline
FE4 & 0.088 & 0.595 & 0.098 & 0.104 & 0.115 & 0.03 \\
\hline
FE5 & 0.103 & 0.140 & 0.151 & 0.157 & 0.449 & 0.06 \\
\hline
\end{tabular}
\end{table}

\begin{figure}[htbp]
\centering
\includegraphics[width=0.9\textwidth, height=0.98\textheight, keepaspectratio]{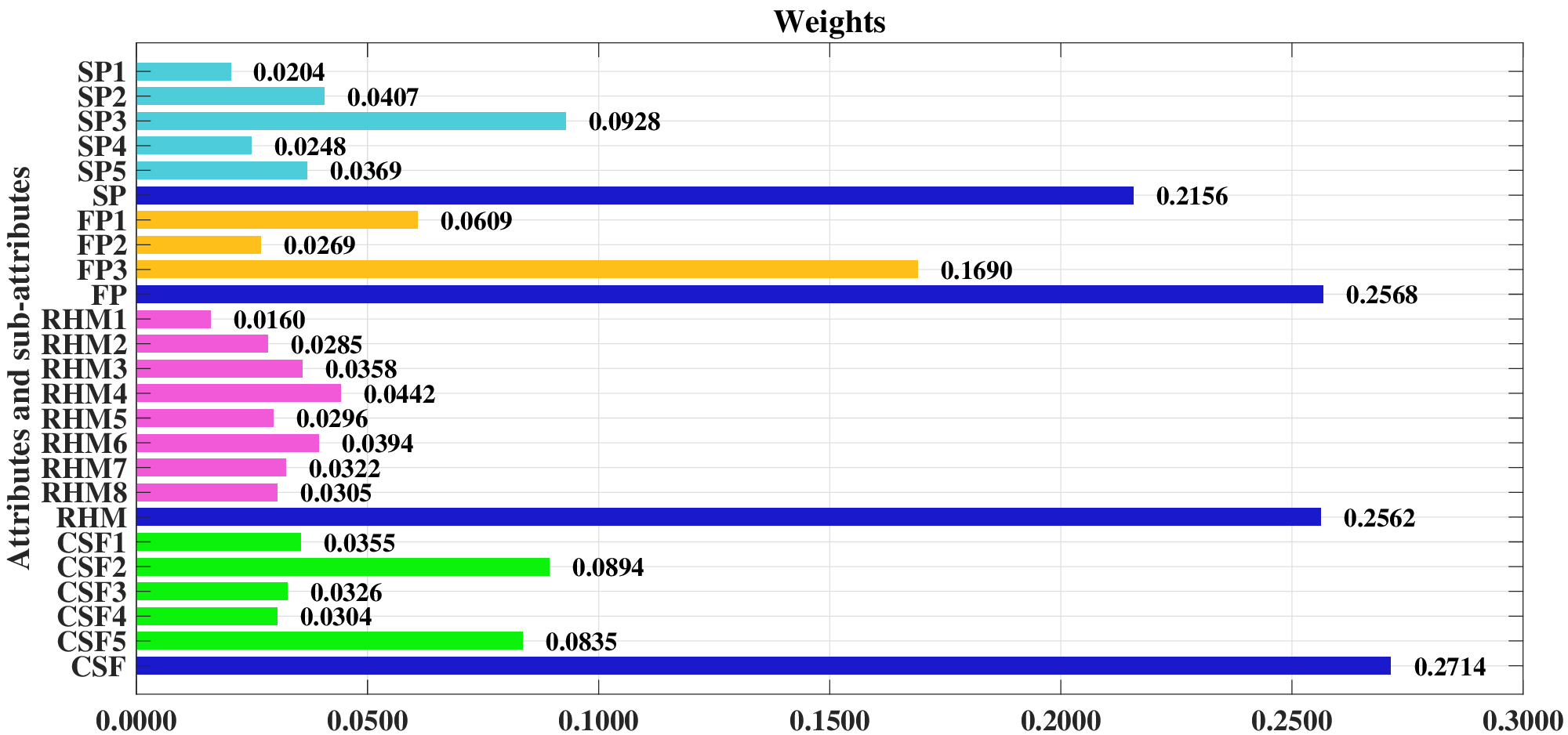}
\caption{Calculated weights of attributes and sub-attributes.}
\label{fig: Calculated weights of attributes and sub-attributes}
\end{figure}

The global weight of each sub-attribute, based on expert input, was calculated by multiplying the weight of its corresponding main attribute (i.e., SP, FP, RHM, and CSF), obtained earlier in this section, with its crisp sub-attribute weight. The final weight of each sub-attribute was then determined by averaging the global weights provided by all experts. Subsequently, the final weight of each main attribute was computed as the sum of the weights of its associated sub-attributes. The resulting weights for both attributes and sub-attributes are illustrated in Figure~\ref{fig: Calculated weights of attributes and sub-attributes}.

Figure~\ref{fig: Calculated weights of attributes and sub-attributes} illustrates that all four main attributes contribute relatively evenly to the fusion siting process, with CSF showing the highest overall importance (27.14\%) and SP the lowest (21.56\%). The experts cumulatively identify FP3 (federal incentives) as the most critical factor in fusion siting among the sub-criteria. Other sub-criteria deemed highly influential include SP3 (energy price), FP1 (net electricity imports), CSF2 (transportation), and CSF5 (substation), with respective weights of 9.28\%, 6.09\%, 8.94\%, and 8.35\%. Conversely, RHM1 (protected lands) receives the least priority from the experts with a weight of 1.60\%. 


\subsection{Determination of the suitability of retired U.S. coal plants (2020-2025) for fusion deployment}

\begin{figure}[htbp]
\centering
\includegraphics[width=.6\textwidth, height=0.6\textheight, keepaspectratio]{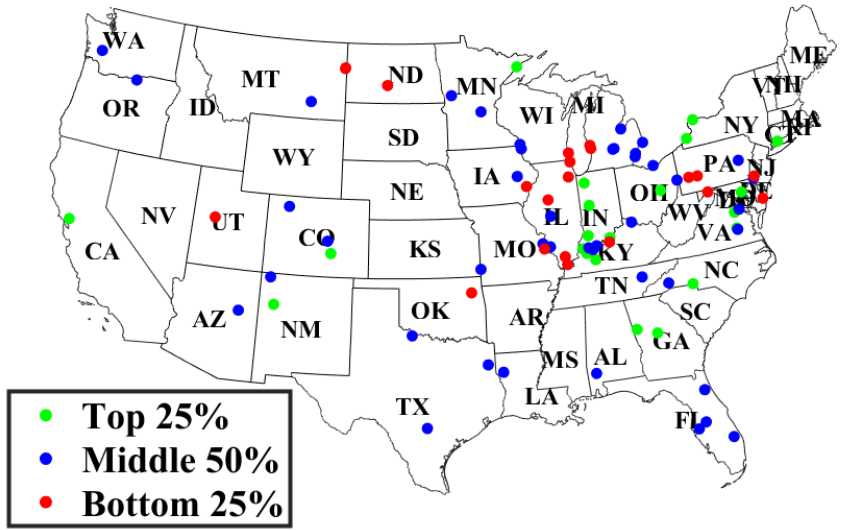}
\caption{Top 25\%, middle 50\%, and bottom 25\% performing coal sites based on all attributes (overall).}
\label{fig: overall attribute}
\end{figure}

After determining the sub-criteria weights, we applied the MARCOS methodology (see Methods section) to rank the 85 retired coal sites in the U.S. between 2020 and 2025. Utility functions (suitability scores) were calculated for all sites (see Code and data availability section), and Table~\ref{tab:top50-coal-sites} presents the plants' name and their corresponding utility values obtained from the MARCOS method up to three decimals. Notably, the “Somerset Operating Co LLC" retired coal plant in New York ranks highest among all sites. It should not be interpreted to indicate that sites ranking higher have always favorable values of the higher weighted sub-attributes (e.g., federal incentives, energy price, net electricity imports, transportation, substation, etc), shown in Figure~\ref{fig: Calculated weights of attributes and sub-attributes}. For instance, the “Somerset Operating Co LLC" coal site benefits from favorable federal incentives and its location in a region with high net-electricity imports. But the site has a reasonably high energy price (21.27 \textcent/kWh, compared to a range of 8.21 \textcent/kWh to 26.83 \textcent/kWh), longer distance to the nearest major road (49,619.79 miles, with a range between 195.89 miles and 176,191.90 miles), and a moderate distance to the nearest substation (153.85 miles, with a range between 40.95 miles to 20,766.58 miles). Figure~\ref{fig: overall attribute} illustrates the top 25\%, middle 50\%, and bottom 25\% performing coal sites while considering all attributes. Threshold values for identifying the top and bottom 25\% performers were determined using the third quartile (Q3) and first quartile (Q1) of the utility score, respectively.

\begin{landscape}
\begin{table}[htbp]
\centering
\small
\caption{List of all retired coal sites between 2022 \& 2025 and their suitability scores for fusion siting.}
\label{tab:top50-coal-sites}

\setlength{\tabcolsep}{3pt}
\renewcommand{\arraystretch}{1.1}

\begin{tabular}{|
    >{\centering\arraybackslash}M{3.4cm}|c|
    >{\centering\arraybackslash}M{3.4cm}|c|
    >{\centering\arraybackslash}M{3.4cm}|c|
    >{\centering\arraybackslash}M{3.4cm}|c|
    >{\centering\arraybackslash}M{3.4cm}|c|}
\hline
\textbf{Rank \& Plant Name} & \(\mathbf{f(K_i)}\) &
\textbf{Rank \& Plant Name} & \(\mathbf{f(K_i)}\) &
\textbf{Rank \& Plant Name} & \(\mathbf{f(K_i)}\) &
\textbf{Rank \& Plant Name} & \(\mathbf{f(K_i)}\) &
\textbf{Rank \& Plant Name} & \(\mathbf{f(K_i)}\) \\
\hline 
1. Somerset (NY)        & 0.637 &
18. Comanche (CO)          & 0.571 &
35. Meramec (MO)         & 0.530 &
52. Dallman (IL)       & 0.489 &
69. Waukegan (IL)        & 0.434 \\

\hline 
2. Dunkirk (NY)              & 0.634 &
19. G G Allen (NC)             & 0.570 &
36. Genoa (WI)             & 0.529 &
53. Erickson (MI)             & 0.488 &
70. Will County (IL)            & 0.430 \\

\hline 
3. Phillips (CA)              & 0.616 &
20. Bridgeport (CT)         & 0.564 &
37. Lansing (IA)         & 0.528 &
54. Lowman (AL)           & 0.483 &
71. Joppa (IL)                & 0.429 \\

\hline 
4. Spruance (VA)               & 0.609 &
21. Herbert (MD)           & 0.557 &
38. C D McIntosh Jr (FL)            & 0.524 &
55. Pirkey (TX)           & 0.480 &
72. Lewis \& Clark (MT)        & 0.427 \\

\hline 
5. R M Schahfer (IN)      & 0.608 &
22. Asheville (NC)              & 0.557 &
39. Indiantown (FL)         & 0.522 &
56. Baldwin (IL)               & 0.480 &
73. Muscatine (IA)     & 0.422 \\

\hline 
6. Wansley (GA)            & 0.603 &
23. F B Culley (IN)        & 0.556 &
40. Avon Lake (OH)         & 0.520 &
57. Trenton (MI)           & 0.477 &
74. Logan (NJ)               & 0.418 \\

\hline 
7. AES Petersburg (IN)    & 0.595 &
24. Cholla (AZ)           & 0.555 &
41. Asbury (MO)           & 0.520 &
58. W H Zimmer (OH)      & 0.475 &
75. Homer City (PA) & 0.414 \\

\hline 
8. Scherer (GA)          & 0.591 &
25. South Plant (CO)        & 0.554 &
42. Boardman (OR)                & 0.520 &
59. Big Bend (FL)              & 0.470 &
76. Cheswick (PA)            & 0.413 \\

\hline 
9. R Gallagher (IN)       & 0.590 &
26. Seminole (FL)         & 0.553 &
43. Elmer Smith (KY)       & 0.519 &
60. TransAlta (WA)             & 0.468 &
77. J B Sims (MI)      & 0.409 \\

\hline 
10. Robert A Reid (KY)    & 0.585 &
27. San Juan (NM)       & 0.553 &
44. River Rouge (MI)               & 0.518 &
61. Eckert (MI)            & 0.466 &
78. E D Edwards (IL) & 0.396 \\

\hline 
11. Birchwood (VA)          & 0.583 &
28. Craig (CO)            & 0.552 &
45. Morgantown (MD)                & 0.514 &
62. Dan E Karn (MI)              & 0.465 &
79. Marion (IL)            & 0.385 \\
 
\hline 
12. A B Brown (IN)        & 0.582 &
29. Hoot Lake (MN)         & 0.547 &
46. Oklaunion (TX)                & 0.513 &
63. Kenneth (KY)              & 0.464 &
80. J H Campbell (MI)            & 0.379 \\

\hline 
13. Paradise (KY)          & 0.578 &
30. Sherburne (MN)         & 0.544 &
47. St Clair (MI)                & 0.507 &
64. Bull Run (TN)              & 0.453 &
81. Mill Creek (KY)            & 0.367 \\

\hline 
14. Taconite (MN)          & 0.577 &
31. Dolet Hills (LA)         & 0.538 &
48. W H Sammis (OH)                & 0.502 &
65. Sidney (MT)              & 0.452 &
82. South Oak (WI)            & 0.366 \\

\hline 
15. Conesville (OH)          & 0.575 &
32. Prairie Creek (IA)         & 0.535 &
49. Colstrip (MT)               & 0.495 &
66. GREC (OK)              & 0.451 &
83. AES Warrior (MD)            & 0.354 \\

\hline 
16. Crawfordsville (IN)         & 0.575 &
33. Chesterfield (VA)         & 0.534 &
50. Chambers (NJ)                & 0.494 &
67. Indian River (DE)              & 0.447 &
84. Intermountain (UT)            & 0.328 \\

\hline 
17. Escalante (NM)         & 0.574 &
34. Wheelabrator (PA)         & 0.533 &
51. J T Deely (TX)                & 0.492 &
68. R M Heskett (ND)              & 0.435 &
85. Rush Island (MO)            & 0.319 \\
\hline
\end{tabular}
\end{table}
\end{landscape}

\begin{figure}[htbp]
    \centering
    
    \begin{subfigure}{0.48\textwidth}
        \includegraphics[width=\linewidth]{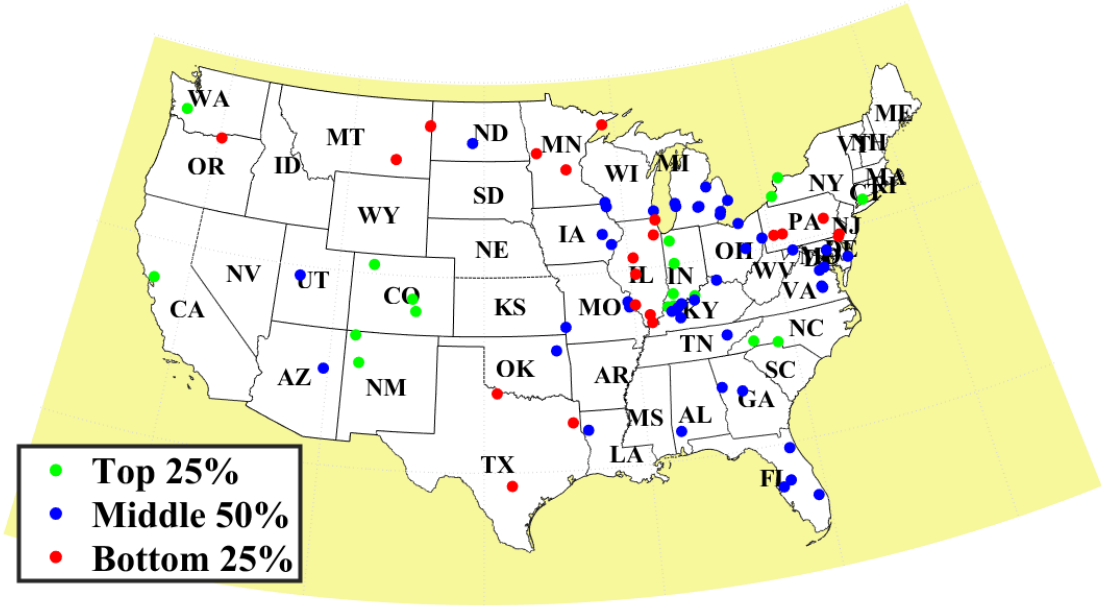}
        \caption{Ranking based on only SP attribute.}
        \label{fig:SP}
    \end{subfigure}
    \hfill
    \begin{subfigure}{0.48\textwidth}
        \includegraphics[width=\linewidth]{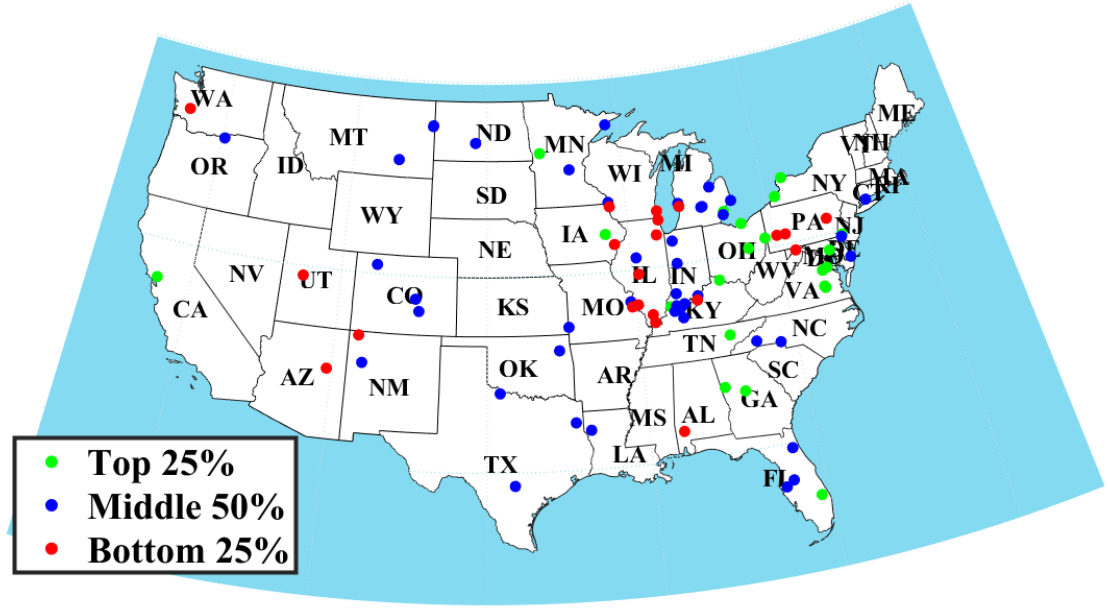}
        \caption{Ranking based on only FP attribute.}
        \label{fig:FP}
    \end{subfigure}
    
    \begin{subfigure}{0.48\textwidth}
        \includegraphics[width=\linewidth]{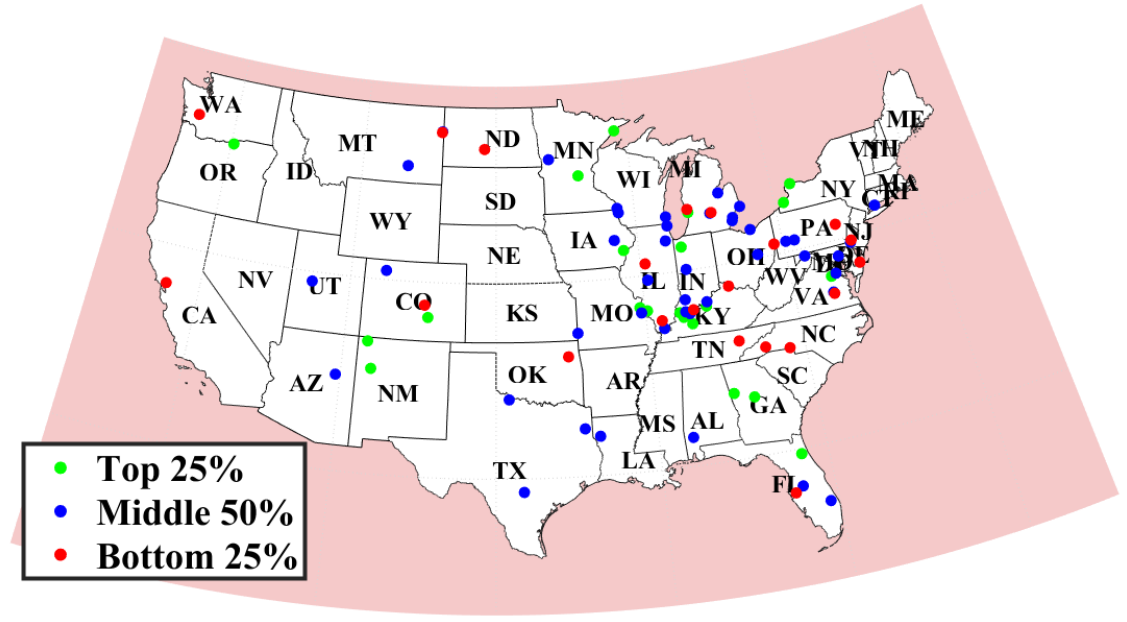}
        \caption{Ranking based on only RHM attribute.}
        \label{fig:RHM}
    \end{subfigure}
    \hfill
    \begin{subfigure}{0.48\textwidth}
        \includegraphics[width=\linewidth]{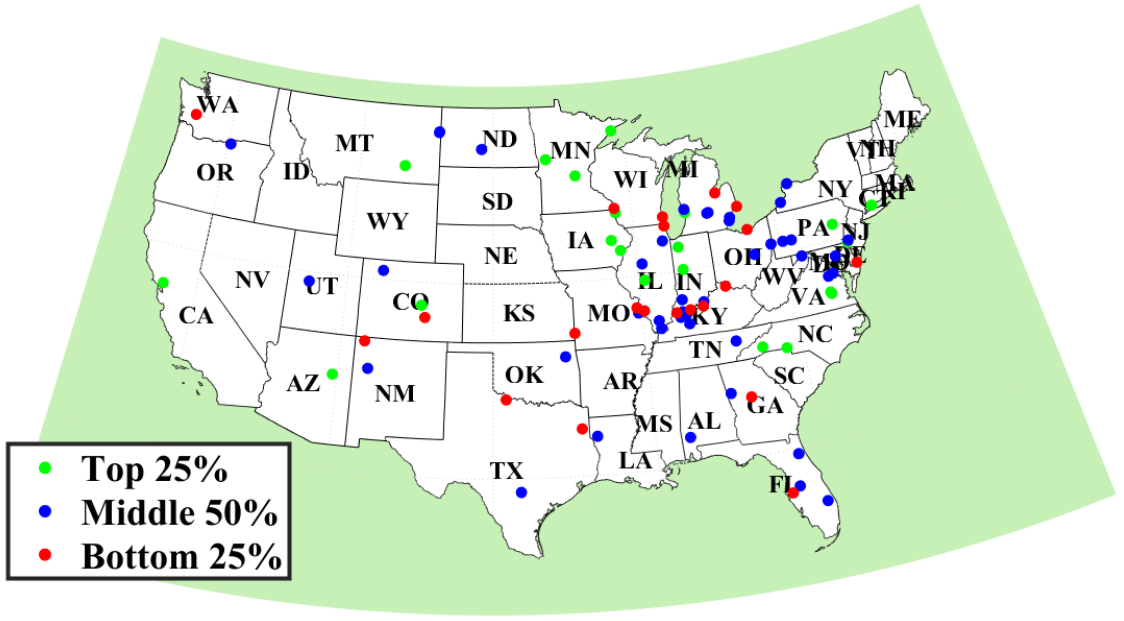}
        \caption{Ranking based on only CSF attribute.}
        \label{fig:CSF}
    \end{subfigure}

    \caption{Top 25\%, middle 50\%, and bottom 25\% performing coal sites based on four attributes (i.e., SP, FP, RHM, and CSF).}
    \label{fig:Combined_US_Map_all_attributes}
\end{figure}

To gain deeper insights, we further analyzed site suitability based on individual attribute categories, such as SP, FP, RHM, and CSF. Figure~\ref{fig:Combined_US_Map_all_attributes} presents the top 25\%, middle 50\%, and bottom 25\% performing retired coal sites based on SP, FP, RHM, and CSF categories. The detailed results, along with the suitability score for each coal site, are provided in the Data and code availability section. The results highlight that “G G Allen (NC)", “Phillips 66 Carbon Plant (CA)", “Boardman (OR)", and “Wheelabrator Frackville Energy (PA)" rank highest in SP, FP, RHM, and CSF, respectively. A key observation from figure~\ref{fig:Combined_US_Map_all_attributes} is that sites highly ranked overall may not necessarily perform best under individual criteria. For instance, “Somerset Operating Co LLC (NY)", the overall top performer site, ranks 14, 11, 5, and 31 in SP, FP, RHM, and CSF criteria, respectively. 
Similarly, a site that leads in one criterion may rank considerably lower in others. For example, although “G G Allen (NC)" ranks top in SP criterion, the site ranks 34, 65, and 17 in FP, RHM, or CSF, respectively. Thus, these figures offer valuable guidance for siting authorities and policymakers seeking to prioritize specific criteria in siting fusion power plants. Our assessment methodology can also inform site comparison and selection on the basis of relative improvements that may need to be made at sites to improve their viability for hosting a fusion power plant.

\subsection{Identification of the sub-attributes that most influence fusion site suitability}

In this section, we assessed the sensitivity of sub-criteria in terms of their alteration potential to cause rank reversals among the studied coal sites. Following the procedure outlined in Methods (see Methods Section), we computed the criticality degree $D'_k$ for each sub-criterion, as summarized in Table~\ref{tab:List of criticality degree values and rank reversal scenarios for all sub-attributes}. A lower value of $D'_k$  indicates a more sensitive or critical sub-criterion, whereas a higher value suggests lower sensitivity. According to Table~\ref{tab:List of criticality degree values and rank reversal scenarios for all sub-attributes}, RHM4 (landslide hazards) emerges as the most sensitive sub-criterion, while RHM3 (fault lines) is the least sensitive. The ranked order of sub-criteria from most to least critical, based on $D'_k$, is as follows: 

Landslide Hazards (RHM4) $<$ Population (CSF1) $<$ Slope (RHM8) $<$ Energy Price (SP3) $<$ Net Electricity Imports (FP1) $<$ 100-Year Flood (RHM6) $<$ Nuclear Restrictions (SP1) $<$ Transportation (CSF2) $<$ Nuclear Inclusive Policies (SP2) $<$ Operating Nuclear Facilities (CSF3) $<$ Construction Labor Rate (SP5) $<$ Nuclear R\&D (CSF4) $<$ Substation (CSF5) $<$  Open Waters and Wetlands (RHM7) $<$ Hazardous Facilities (RHM2) $<$ Market Regulation (SP4) $<$ Hydrogen Demand (FP2) $<$ Federal Incentives (FP3) $<$ Safe Shutdown Earthquake (RHM5) $<$ Protected Lands (RHM1) $<$ Fault Lines (RHM3).

Table~\ref{tab:List of criticality degree values and rank reversal scenarios for all sub-attributes} presents the sub-attributes and their corresponding criticality degree $D'_k$  values. A smaller $D'_k$ implies that even a minor percentage change in a sub-criterion could alter the ranking of sites. For instance, a 0.1512\% change in nuclear restrictions (SP1) results in a rank reversal. While a 0.1\% variation is generally significant, such a change in nuclear restrictions (SP1) is relatively usual due to its binary nature. We obtained very small critical degree values for several cases in Table~\ref{tab:List of criticality degree values and rank reversal scenarios for all sub-attributes} due to a large number of sub-attributes, roughly equal weights for several sub-criteria, and several sub-criteria having binary values. 
Across the twenty-one sub-criteria, we observed substantial variation in both data type and numeric scale. Nine sub-criteria, i.e., nuclear inclusive policies, market regulation, federal incentives, fault lines, landslide hazards, safe shutdown earthquake, 100-year flood, open waters and wetlands, and slope, are binary variables, capturing the presence or absence of each condition. Several other sub-criteria have relatively small integer ranges, such as nuclear restrictions (0–2), protected lands (0–2), hazardous facilities (0–16), and nuclear R\&D (0–4).
In contrast, the remaining sub-criteria exhibited much larger numerical magnitudes. For example, energy price spans 8.21–26.83 cents/kWh, construction labor rates range from USD 33,580–63,186, net electricity imports vary between –80,245 million kWh and 74,707 million kWh, and hydrogen demand ranges from 0–269.43 kt/yr. Similar wide ranges are observed in population (299.74–226,829.8 miles), transportation distance (195.89–176,191.90 miles), proximity to operating nuclear facilities (5.53–735.37 miles), and distance to substations (40.95–20,766.58 miles).
This wide heterogeneity in both data type (binary vs. continuous) and magnitude, coupled with the large number of sub-criteria, leads to comparatively small values in the computed criticality degree.

\begin{table}[htbp]
\centering
\small
\caption{List of sub-attributes and their corresponding criticality degree values.}
\label{tab:List of criticality degree values and rank reversal scenarios for all sub-attributes}
\begin{tabular}{|c|c|c|}
\hline
\textbf{Sub-attributes} & \( D'_k \) \\
\hline
Nuclear Restrictions (SP1) & 0.1512 \\
\hline
Nuclear Inclusive Policies (SP2) & 0.2833 \\
\hline
Energy Price (SP3) & 0.1182 \\
\hline
Market Regulation (SP4) & 0.9299 \\
\hline
Construction Labor Rate (SP5) & 0.3777 \\
\hline
Net Electricity Imports (FP1) & 0.1485 \\
\hline
Hydrogen Demand (FP2) & 1.3107 \\
\hline
Federal Incentives (FP3) & 2.5086 \\
\hline
Protected Lands (RHM1) & 3.8699 \\
\hline
Hazardous Facilities (RHM2) & 0.4698 \\
\hline
Fault Lines (RHM3) & 25.6364 \\
\hline
Landslide Hazards (RHM4) & 0.0697 \\
\hline
Safe Shutdown Earthquake (RHM5) & 3.5655 \\
\hline
100-Year Flood (RHM6) & 0.1506 \\
\hline
Open Waters and Wetlands (RHM7) & 0.4149 \\
\hline
Slope (RHM8) & 0.1010 \\
\hline
Population (CSF1) & 0.0977 \\
\hline
Transportation (CSF2) & 0.2482 \\
\hline
Operating Nuclear Facilities (CSF3) & 0.3690 \\
\hline
Nuclear R\&D (CSF4) & 0.3904 \\
\hline
Substation (CSF5) & 0.3935 \\
\hline
\end{tabular}
\end{table}

\subsection{Investigation on how industry-selected coal sites align with our fusion siting assessment}

In this section, we examined three U.S. retired coal sites that have recently attracted significant attention from leading fusion companies for potential fusion power deployment.
Type One Energy signed a cooperative agreement with the utility Tennessee Valley Authority (TVA) to repurpose the retired Bull Run coal plant into a fusion power plant with their “Infinity” stellarator design.
The retired Bull Run plant (coal-fired, decommissioned in 2023) is specifically named in media coverage as the planned fusion site. The planned fusion plant is referred to as “Infinity Two” (a 350 MWe stellarator) targeted for the mid-2030s~\cite{CATF2025}.
Zap Energy is conducting a feasibility study for a first-of-its-kind Z-pinch fusion pilot plant sited at the retired coal sites in Centralia, Washington~\cite{ZAP2022}.
Commonwealth Fusion Systems (CFS) has planed to build its first commercial fusion power plant, ARC, on a site owned by Dominion Energy near a retired coal plant in Chesterfield County, Virginia~\cite{CFSVirginia}. While public-facing media sometimes describes the site as being “near a retiring coal plant,” we found no publicly available source that clearly states the ARC project site is the former coal plant site itself (i.e., that it’s built on the coal plant’s footprint, rather than “nearby”). Therefore, we considered the “Chesterfield (VA)" retired coal site, along with the “Bull Run (TN)" and “Centralia Generation (WA)" retired coal sites, here for comparative analysis. Table~\ref{tab: Ranking of the fusion deployment sites considered by Type One Energy, Zap Energy, and Commonwealth Fusion Systems} lists these three retired coal sites and their corresponding ranks obtained from our methodology and results.

\begin{table}[htbp]
\centering
\small
\caption{Ranking of the fusion deployment sites considered by Type One Energy, Zap Energy, and Commonwealth Fusion Systems.}
\label{tab: Ranking of the fusion deployment sites considered by Type One Energy, Zap Energy, and Commonwealth Fusion Systems}
\begin{tabular}{|M{3.2cm}|M{2.6cm}|M{1.8cm}|M{1.8cm}|M{2.0cm}|M{2.0cm}|}
\hline
\multirow{2}{*}{\textbf{Plant name}} & \multicolumn{5}{c|}{\textbf{Ranking based on:}} \\
\cline{2-6}
 & \textbf{All attributes} & \textbf{Only SP attribute} & \textbf{Only FP attribute} & \textbf{Only RHM attribute} & \textbf{Only CSF attribute} \\
\hline
Bull Run (TN) &  64 &  31 &  9 &  83 & 52 \\
\hline
Centralia (WA) &  60 &  18 &  68 &  70 & 79 \\
\hline
Chesterfield (VA) & 33 &  33 &  2 &  75 & 16 \\
\hline
\end{tabular}
\end{table}

From Table~\ref{tab: Ranking of the fusion deployment sites considered by Type One Energy, Zap Energy, and Commonwealth Fusion Systems} and Figure~\ref{fig:Combined_US_Map_all_attributes}, three key patterns appear for the candidate sites selected by Type One Energy, Zap Energy, and Commonwealth Fusion Systems.
“Bull Run (TN)" exhibits mid-level suitability when considering all attributes collectively, as well as when evaluated solely on SP and CSF attributes. However, it performs strongly under the FP category and poorly under the RHM criteria.
“Centralia Generation (WA)", the site associated with Zap Energy, also ranks as a mid-level performer overall. It stands out as a top-tier site when assessed only on SP factors, but it consistently falls into the bottom tier across FP, RHM, and CSF categories.
“Chesterfield (VA)", associated with Commonwealth Fusion Systems, is also a mid-level performer overall and under SP attributes, but it ranks as a top-tier site under FP and CSF attributes while dropping to the bottom tier for RHM considerations.
To better understand the drivers behind these variations, we further conducted a sub-attribute–level assessment for all three sites. The detailed values for each sub-attribute are presented in Table~\ref{tab: Comparison of the sub-attributes for the three candidate sites considered by Type One Energy, Zap Energy, and Commonwealth Fusion Systems}, enabling a granular comparison of the factors shaping each site's suitability profile.

\begin{table}[htbp]
\centering
\small
\caption{Comparison of the sub-attributes for the three candidate sites considered by Type One Energy, Zap Energy, and Commonwealth Fusion Systems.}
\label{tab: Comparison of the sub-attributes for the three candidate sites considered by Type One Energy, Zap Energy, and Commonwealth Fusion Systems}
\begin{tabular}{|M{6.2cm}|M{2.5cm}|M{2.5cm}|M{2.5cm}|}
\hline
\multirow{2}{*}{\textbf{Sub-attributes}} & \multicolumn{3}{c|}{\textbf{Plant name}} \\ \cline{2-4}
 & Bull Run (TN) & Centralia (WA) & Chesterfield (VA) \\ 
\hline
Nuclear Restrictions (SP1) [Unit: Count] & 0 & 0 & 0\\
\hline
 Nuclear Inclusive Policies (SP2) [Unit: Binary (Yes/No)] & No & Yes & No\\
\hline
 Energy Price (SP3) [Unit: Cents/kWh] & 11. 73 & 10.88 & 11.27\\
\hline
 Market Regulation (SP4) [Unit: Binary (Yes/No)] & Yes & Yes & Yes\\
\hline
 Construction Labor Rate (SP5) [Unit: USD (5-yr avrg.)] & 32,270 & 54,658 & 35,824\\
\hline
 Net Electricity Imports (FP1) [Unit: Million kWh] & 31,615 & -20,290 & 52,123\\
\hline
 Hydrogen Demand (FP2) [Unit: kt/yr] & 0 & 0 & 37.07\\
\hline
 Federal Incentives (FP3) [Unit: Binary (Yes/No)] & Yes & Yes & Yes\\
\hline
 Protected Lands (RHM1) [Unit: Count] & 0 & 0 & 0\\
\hline
 Hazardous Facilities (RHM2) [Unit: Count] & 2 & 2 & 4\\
\hline
 Fault Lines (RHM3) [Unit: Binary (Yes/No)] & No & No & No\\
\hline
 Landslide Hazards (RHM4) [Unit: Binary (Yes/No)] & Yes & Yes & Yes\\
\hline
 Safe Shutdown Earthquake (RHM5) [Unit: Binary (Yes/No)] & Yes & No & Yes\\
\hline
 100-Year Flood (RHM6) [Unit: Binary (Yes/No)] & Yes & No & No\\
\hline
 Open Waters and Wetlands (RHM7) [Unit: Binary (Yes/No)] & No & No & Yes\\
\hline
 Slope (RHM8) [Unit: Binary (Yes/No)] & No & Yes & Yes\\
\hline
 Population (CSF1) [Unit: Miles]  & 11,346.42 & 45,869.50 & 23,213.10\\
\hline
Transportation (CSF2) [Unit: Miles]  & 12,755.69 & 13,219.71 & 2,752.52\\
\hline
Operating Nuclear Facilities (CSF3) [Unit: Miles]  & 56.51 & 245.57 & 50.95\\
\hline
Nuclear R\&D (CSF4) [Unit: Count]   & 1 & 0 & 0\\
\hline
Substation (CSF5) [Unit: Miles]   & 294.68 & 568.05 & 95.86\\
\hline
\end{tabular}
\end{table}

Table~\ref{tab: Comparison of the sub-attributes for the three candidate sites considered by Type One Energy, Zap Energy, and Commonwealth Fusion Systems} shows that all three sites receive identical scores for nuclear restrictions (SP1). However, “Centralia Generation (WA)" benefits uniquely from nuclear-inclusive (fission) policies (SP2), whereas the other two sites do not. Given the relatively high weight of SP2 (4.07\%), this sub-attribute meaningfully improves Centralia’s overall suitability.
For energy price (SP3), the values across all three sites are nearly identical. Similarly, each site is located within a regulated electricity market, resulting in comparable performance under the market regulation (SP4) sub-attribute.
In contrast, the construction labor rate (SP5) varies more notably. “Bull Run (TN)" and “Chesterfield (VA)" exhibit similar and comparatively lower labor costs than “Centralia Generation (WA)". Because SP5 carries a moderate weight (3.69\%), these lower labor rates give “Bull Run (TN)" and “Chesterfield (VA)" a relative advantage under this sub-attribute.

“Centralia Generation (WA)” is a net electricity exporter (indicated by the negative sign), whereas both “Bull Run (TN)” and “Chesterfield (VA)” are net electricity importers. Because states or sites with higher net electricity imports are more likely to prioritize new technologies to meet demand and improve energy self-sufficiency, “Chesterfield (VA)” benefits substantially from the net electricity import sub-attribute (FP1). This advantage is further strengthened by the relatively high criteria weight assigned to FP1 (6.09\%), making “Chesterfield (VA)” more attractive for fusion deployment than the other two locations.
“Chesterfield (VA)” is the only site with non-zero hydrogen demand (FP2), which further enhances its suitability for fusion deployment, given fusion’s potential role in clean hydrogen production. In contrast, both “Bull Run (TN)” and “Centralia Generation (WA)” exhibit zero hydrogen demand, providing no additional benefit under this criterion.
All three sites qualify as “Energy Communities” under federal policy because all coal-fired generating units at each location have been fully retired. Hence, they are equally advantaged with respect to federal incentives (FP3).

All three sites receive identical scores for the protected lands (RHM1) sub-attribute because none are located near designated protected areas. In contrast, “Chesterfield (VA)” is situated within five miles of four hazardous facilities, whereas “Bull Run (TN)” and “Centralia Generation (WA)” each have two such facilities nearby. As a result, the latter two sites are more favorable under the hazardous facilities (RHM2) sub-attribute. However, its influence on the overall ranking is limited due to its relatively low weight (2.85\%).
Despite the higher weight assigned to fault lines (RHM3, 3.58\%), all three locations perform equally well because none are located near known seismic faults. Similarly, the landslide hazard (RHM4) sub-attribute provides an identical effect across the three sites. However, “Centralia Generation (WA)” is located in an area with peak ground acceleration above 0.3g, meaning it does not benefit from the safe shutdown earthquake (RHM5) sub-criterion, whereas “Bull Run (TN)” and “Chesterfield (VA)” both receive benefits from RHM5.
Regarding flood risk, “Bull Run (TN)” has a 1\% annual chance of flooding, while the other two locations do not. Thus, “Centralia Generation (WA)” and “Chesterfield (VA)” score more favorably under the 100-year flood (RHM6) sub-attribute, which has a reasonably high weight of 3.94\%. For open waters and wetlands (RHM7) sub-attribute, “Chesterfield (VA)” is located closer to open waters and wetlands, making “Bull Run (TN)” and “Centralia Generation (WA)” comparatively better positioned for this criterion (weighted at 3.22\%). Both “Centralia Generation (WA)” and “Chesterfield (VA)” exhibit slopes below the 12\% threshold assumed in this study~\cite{abdussami2024investigation}, resulting in higher suitability under the slope (RHM8) sub-attribute relative to “Bull Run (TN).”

Among the three sites, “Centralia Generation (WA)" has the greatest distance to a population center of more than 25,000 residents, while “Bull Run (TN)" has the shortest. As a result, “Centralia Generation (WA)" gains the most advantage from the population (CSF1) sub-attribute, which has a moderate influence on the ranking (weight: 3.55\%). In contrast, “Chesterfield (VA)" is located very close to major roads, giving it a substantial advantage under the transportation (CSF2) sub-attribute, one of the most influential criteria with a weight of 8.90\%. “Chesterfield (VA)" also benefits from the shortest distance to an existing operating nuclear facility, although “Bull Run (TN)" performs similarly on this metric.
For Nuclear R\&D (CSF4), only “Bull Run (TN)" is located within 100 miles of a nuclear R\&D facility, giving it a moderate benefit due to the sub-attribute’s weight of 3.04\%. Meanwhile, “Chesterfield (VA)" is located very close to a major electric substation. This proximity provides a strong advantage under the substation (CSF5) sub-attribute, which is highly weighted (8.35\%).
Taken together, the comparison in Table~\ref{tab: Ranking of the fusion deployment sites considered by Type One Energy, Zap Energy, and Commonwealth Fusion Systems} and Table~\ref{tab: Comparison of the sub-attributes for the three candidate sites considered by Type One Energy, Zap Energy, and Commonwealth Fusion Systems} indicates that “Chesterfield (VA)" could be a more suitable site for fusion power siting compared to the other two, given that weights of the sub-attributes are assumed as per Figure~\ref{fig: Calculated weights of attributes and sub-attributes}. This outcome is primarily driven by Chesterfield’s significant advantages in net electricity imports (FP1), transportation access (CSF2), and substation proximity (CSF5).  Notably, transportation access and substation proximity are both factors that can be improved for the other two sites and could be prioritized by the two companies focusing on these sites.

\section{DISCUSSIONS AND CONCLUSIONS}


This study presents a comprehensive approach to siting fusion facilities by evaluating a diverse set of 21 sub-criteria categorized under four primary attributes: SP, FP, RHM, and CSF. By integrating expert input with advanced MCDM methods, the analysis identifies the relative importance of each factor in the siting process. Although traditional FUCOM could be applied due to the availability of specific values, we adopted F-FUCOM to better address the uncertainty and linguistic vagueness inherent in the experts’ opinions. The results demonstrate a balanced contribution from all four main attributes, with CSF having the highest influence (27.14\%) and SP being the lowest (21.56\%). Among the sub-criteria, federal incentives (FP3) emerge as the most critical factor, followed by energy price (SP3), transportation infrastructure (CSF2), substation availability (CSF5), and net electricity imports (FP1). By contrast, protected lands (RHM1) is assigned the lowest significance, based on our expert elicitation, although in practice this criterion may carry greater importance, as regulatory, environmental, and social constraints can make such areas highly restrictive for project deployment. These findings highlight the multifaceted nature of fusion siting and underscore the need for holistic planning that integrates policy, infrastructure, and community readiness.

Federal incentives (FP3) ranks the highest because fusion energy is an emerging and capital-intensive technology. Governmental support through tax credits, grants, and regulatory streamlining can significantly impact the success of early FOAK fusion deployments. Experts likely view these incentives as critical enablers for market entry and risk reduction. 
Energy price (SP3) is a proxy for market competitiveness. High energy prices may create stronger incentives for alternative energy sources, such as fusion, while low prices could deter investment. States with high retail or wholesale prices may offer better revenue potential for fusion developers. 
Fusion facilities will require significant infrastructure for component delivery (e.g., superconducting magnets, cryogenic systems), construction materials, and ongoing operational logistics. Proximity to highways, rail, and ports reduces logistical costs and complexity, especially for large FOAK components that are not modularized. Hence, transportation (CSF2) is significant in the siting of fusion plants.

A fusion plant requires electricity from the grid for plasma heating and must be able to transmit its generated electricity to the grid efficiently. Existing substations enable grid interconnection without extensive new infrastructure, making sites with substations more attractive for rapid deployment. Thus, substation (CSF5) also plays a vital role in fusion siting. 
Regions with high electricity imports (FP1) often face issues related to energy security or supply adequacy. Fusion plants located in these regions could replace imported energy with domestically produced, carbon-free baseload power, thereby improving reliability and economic self-sufficiency.
On the other hand, protected lands (RHM1) are legally excluded from energy development, so their importance is more about exclusion rather than selection. Experts may have considered this a “screening criterion” rather than a key differentiator among candidate sites.

In the case study, the research identified “Somerset Operating Co LLC (NY)” as the most suitable site for fusion deployment when considering all criteria collectively. However, individual attribute analysis revealed that different sites lead in specific categories—“G G Allen (NC)” in SP, “Phillips 66 Carbon Plant (CA)” in FP, “Boardman (OR)” in RHM, and “Wheelabrator Frackville Energy (PA)” in CSF. This highlights the importance of a multi-criteria approach, as top-ranked sites overall may not consistently perform best across individual factors. It is inappropriate to draw definitive conclusions based solely on a few top-weighted factors. The overall ranking depends on the combination of all attributes, the accuracy of their data, and their assigned weights. While we also intended to identify potential “non-starter” sites—those that consistently rank in the bottom 25\% across all four attribute categories—our analysis revealed that no site consistently fell into the lowest quartile across all dimensions. This result suggests that even the lowest-ranked sites in one attribute group may possess redeeming qualities in others, reinforcing the value of a multi-attribute and balanced siting framework. This finding carries important implications. First, it demonstrates that no site should be excluded outright without a comprehensive evaluation. Second, it emphasizes that site feasibility is not binary; rather, improvements in a subset of policy or infrastructure conditions (e.g., federal or state incentives) could significantly elevate a site's overall viability. Finally, this outcome supports flexibility and adaptability in fusion siting strategy, allowing stakeholders—state and federal agencies, utilities, and communities—to make targeted interventions to upgrade underperforming locations, rather than abandoning them entirely.

The sensitivity analysis reveals that landslide hazards (RHM4) is the most critical sub-criterion for this fusion siting framework. In contrast, fault lines (RHM3) is the least sensitive. The comparative analysis of the three candidate sites, i.e., “Bull Run (TN)", “Centralia Generation (WA)", and “Chesterfield (VA)", chosen by three different fusion industries, reveals that “Chesterfield (VA)" demonstrates the strongest overall suitability for fusion deployment when weighted sub-attributes are considered. While several criteria, such as nuclear restrictions, energy prices, market regulation, fault lines, and landslide hazards, affect all three sites similarly, “Chesterfield (VA)" gains significant advantages from high-impact sub-attributes, particularly its substantial net electricity imports, excellent transportation access, and close proximity to a major electric substation. These factors carry some of the highest weights in the framework and collectively elevate Chesterfield’s ranking. In contrast, “Centralia Generation (WA)" benefits from nuclear inclusive policies, while “Bull Run (TN)" gains moderate advantages from nuclear R\&D proximity and construction labor costs. However, these strengths are associated with lower-weighted sub-attributes and do not offset Chesterfield’s advantages.

The proposed fusion siting methodology offers a transparent and adaptable framework that can benefit a wide range of stakeholders, from federal and state agencies to utilities, local governments, and communities. By decomposing the decision-making process into clearly defined attributes and sub-attributes with expert-assigned weights, the study allows community members and regional planners to understand exactly how sites are being evaluated. This openness can foster public trust and transparency, a crucial element for gaining local acceptance for fusion deployment. Importantly, the framework enables community-driven feasibility analysis: individuals or local organizations can re-rank or screen sites based on the attributes they consider most relevant or based on site-specific knowledge. For instance, a community interested in attracting fusion deployment can use the framework to evaluate how their site fares and which factors they might influence to improve their site' viability.

Moreover, the methodology distinguishes between unchangeable site characteristics (e.g., seismicity, flood risk) and policy or infrastructure-related factors that are subject to intervention, such as nuclear-inclusive state policies, market regulation, federal incentives, or regional infrastructure like substations or R\&D centers. This distinction is vital for decision-makers: while they may not be able to alter a site's natural hazards, they can implement policy reforms, incentive programs, or infrastructure investments to elevate the suitability of otherwise underperforming sites.

The results also inform the strategic placement of regional and federal incentives. Knowing which attributes contribute most to site viability allows governments to direct resources where they will be most effective, either by bolstering enabling conditions in promising regions or by supporting necessary reforms in marginal areas. From a utility perspective, the study provides a clear signal of which regulatory or infrastructural changes might unlock new siting options. Overall, this methodology empowers multiple layers of stakeholders: Federal agencies can tailor incentives and national siting frameworks; states can modify enabling policies to attract fusion investments; utilities can prioritize sites based on regulatory and connectivity advantages; local governments gain clarity on where and how to advocate for site readiness; and communities are invited into the process, with access to understandable, evidence-based tools that support digital democracy and informed consent for future fusion development.

For future work, we plan to develop a user-friendly decision-support tool that incorporates additional siting attributes, stakeholder inputs, and customizable weighting schemes. Such a tool will enable policymakers and industry stakeholders to assess site suitability more flexibly and transparently.
This study does not account for how site selection might affect the transportation and management of fusion-related waste (especially the large volume of low-level waste), which could become an important consideration as deployment scales. Additionally, due to data limitations, we do not incorporate public sentiment on fusion energy into our analysis, despite its growing relevance in the siting of energy infrastructure. Future research should integrate community perspectives and social acceptance metrics of fusion energy to ensure more inclusive and politically feasible fusion deployment strategies.
Moreover, this study adopts an n-th of fusion reactor deployment assumption while applying it to selected fusion power sites that represent FOAK opportunities, highlighting the need for future FOAK-specific siting analyses. This future work will account for the unique risks, infrastructure needs, and uncertainties associated with early-stage fusion deployment, offering more realistic insights into initial siting challenges.

\newpage

\newpage

\subsection*{Data and code availability}


All original data reported in this paper are available at \url{https://github.com/rafisami/Fusion_Siting_MRA_AV}. Any additional information required to reanalyze the data reported in this paper is available from the lead contact upon request.


\section*{ACKNOWLEDGMENTS}


This work is sponsored by the Department of Energy Office of Nuclear Energy under project number (DE-NE0009382), which is funded through the Nuclear Energy University Program (NEUP). We are also grateful to the five fusion experts for their valuable insights and participation in the survey and interview process.

\section*{AUTHOR CONTRIBUTIONS}


Conceptualization, M.R.A. and A.V.; methodology, M.R.A. and A.V.; investigation, M.R.A. and A.V.; writing-–original draft, M.R.A.; writing-–review \& editing, M.R.A. and A.V.; funding acquisition, A.V.; resources, M.R.A., K.D., G.H., and A.V.; supervision, M.R.A. and A.V.

\section*{DECLARATION OF INTERESTS}


The authors declare no competing interests.









\newpage

\bibliography{references}

\bigskip


\end{document}